# Exploring Trade Openness and Logistics Efficiency in the G20 Economies: A Bootstrap ARDL Analysis of Growth Dynamics

Haibo Wang[1[0000-0002-8580-829X]], Lutfu S.Sua[2[0000-0003-4395-865X]]

[1]*Division of International Business and Technology Studies, A.R. Sánchez Jr. School of Business, Texas A&M International University, Laredo, TX, USA, hwang@tamiu.edu*

[2]*Department of Management and Marketing, Southern University and A&M College, Baton Rouge, LA, USA, lutfu.sagbansua@sus.edu*

**Abstract**

This study examines the relationships among trade openness, logistics performance, and economic growth across G20 economies. Using a Bootstrap Autoregressive Distributed Lag (ARDL) model augmented by a dynamic error correction mechanism (ECM), the analysis quantifies the short- and long-run effects of trade facilitation and logistics infrastructure, as measured by the World Bank's Logistics Performance Index (LPI), on economic growth from 2007 to 2023. The G20, as a consortium of the world's leading economies, exhibits significant variation in logistics efficiency and degrees of trade openness, providing a robust context for comparative analysis. The ARDL-ECM approach, reinforced by bootstrap resampling, delivers reliable estimates even with small samples and complex variable linkages. Findings are intended to inform policymakers seeking to enhance trade competitiveness and economic development through targeted investment in infrastructure and regulatory reforms supporting trade facilitation. The results underscore the critical role of efficient logistics—specifically customs administration, physical infrastructure, and shipment reliability—in driving international trade and fostering sustained economic growth. Improvements in these areas can increase a country's trade capacity and overall economic performance.

**Keywords:** ARDL test, Trade Openness, Logistics Efficiency, Bootstrap technique, Dynamic ECM

**Declaration of Competing Interest**

The authors declare that they have no known competing financial interests or personal relationships that could have influenced the work reported in this paper.

**Declaration of Funding**

No funding was received to assist with the preparation of this manuscript.

**Declaration of Ethics**

This study did not involve Human Participants and/or Animals

**Declaration of Data and Code Availability**

Data and Code of this study are available upon request.

**JEL** F32 · F14 · C22 · C32 · N55 · Q43

# 1. Introduction

In an increasingly interconnected global economy, the efficient facilitation of cross-border trade in goods and services has become a fundamental determinant of economic growth and national prosperity. Countries with advanced logistics infrastructure and a high degree of trade openness tend to experience accelerated economic expansion, improved international competitiveness, and elevated living standards. However, the relationships among trade openness, logistics performance, and economic growth are complex, particularly within the G20 group, where structural economic differences pose unique challenges and dynamics. The G20 economies, which account for over 75% of global trade and 80% of world GDP (Huang et al., 2023), play a pivotal role in shaping international trade flows. Despite their collective dominance, these nations exhibit significant heterogeneity in trade facilitation policies, the quality of logistics infrastructure, and levels of global market integration. This diversity provides a valuable context for examining how improvements in trade openness and logistics performance can influence economic growth trajectories within the G20.

Logistics efficiency—encompassing trade infrastructure quality, customs procedures, and shipment reliability—has become a key determinant of international economic competitiveness. The World Bank's Logistics Performance Index (LPI) serves as a comprehensive metric, evaluating countries on dimensions such as customs efficiency, infrastructure quality, ease of arranging shipments, logistics services competence, tracking and tracing capabilities, and the timeliness of deliveries. Trade openness, typically measured by the ratio of exports and imports to GDP, indicates the extent of a country's integration into the global marketplace. A robust analysis of these factors is essential for policymakers seeking to enhance economic growth and international competitiveness, as improvements in logistics and increased trade openness are linked to lower trade costs, expanded market access, and stronger, sustained economic growth. Empirical studies have reinforced these linkages. For example, Marti et al. (2014) employed a gravity model to demonstrate that improved LPI components significantly boost trade flows, particularly in emerging markets such as Africa, South America, and Eastern Europe. Rezaei et al. (2020) used the Best-Worst Method to identify infrastructure as the most influential LPI subcomponent. Liu et al. (2018) examined the relationship between LPI scores and carbon emissions in Asia, illustrating the broader environmental implications of logistics performance. Evaluations by Polat et al. (2023) employed data envelopment analysis (DEA) to assess logistics efficiency across OECD and global economies, whereas Mesic et al. (2022) utilized multi-criteria methods to rank Western Balkan countries by LPI subcomponents. Such methodologies inform best practices for evaluating logistics and trade facilitation, providing relevant frameworks for comparative analysis of G20 economies.

The G20 economies play a crucial role in shaping global economic momentum, with several members exhibiting strong economic performance and increasing trade activity. As major actors in the international marketplace, these countries will influence global trade policies and international negotiations. Substantial investments are critical in logistics infrastructure, including ports, airports, and transportation systems, to support trade and economic growth. However, the G20 economies are facing trade frictions and increasing protectionist pressures, such as recent tariff escalations between the United States and other leading economies, which could have reshaped global trade flows and economic performance. The G20 economies must also address infrastructure constraints, including congestion at key logistics hubs like ports and airports, as well as gaps and inefficiencies in transportation networks. Furthermore, they must balance their economic growth with environmental and social concerns, including reducing carbon emissions and improving labor standards in the logistics sector.

To examine these interconnected factors, this study implements a Bootstrap Autoregressive Distributed Lag (ARDL) approach with a dynamic error correction model (ECM) to assess the short-term and long-term impacts of logistics efficiency and trade openness on economic growth in G20 economies between 2007 and 2023. In general, Autoregressive models, including ARDL, require a large sample size to produce reliable estimates for time series. However, the bootstrap techniques implemented by this study can help mitigate the effects of small sample sizes**.** The conceptual structure guiding this study is presented in Figure 1, followed by research questions for the indicators extracted from World Bank Open Data.

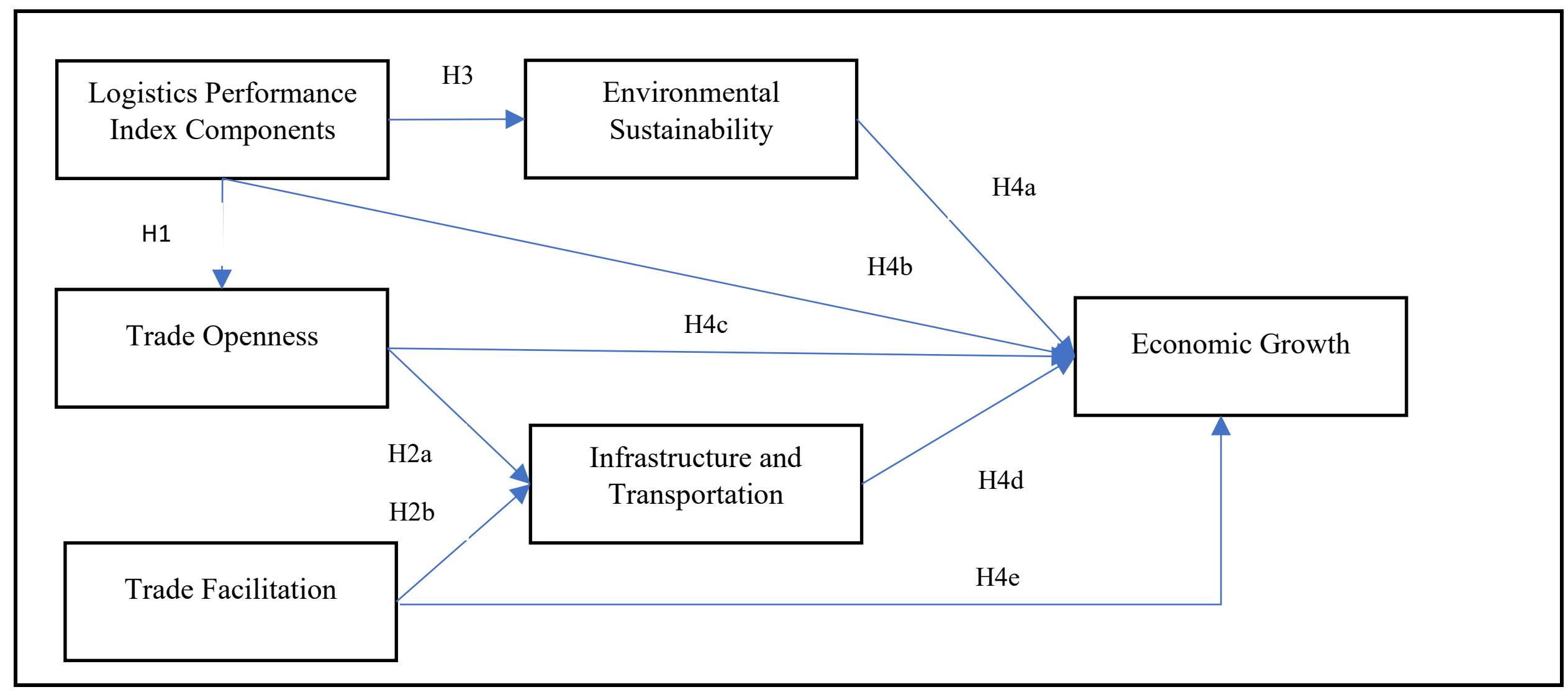

Figure 1. Conceptual framework

RQ1. How does the *logistics performance index* (LPI), including *customs efficiency, infrastructure quality,* and *timeliness*, affect *trade openness* in G20 economies?

H1. The components of the *logistics performance index (LPI)* consistently influence *trade openness* among G20 economies over the long term.

H1a. *Trade openness* in G20 economies is significantly correlated with the long-term trends of the *LPI overall score component (lpi1)*.

H1b. The *LPI customs efficiency component (lpi2*) is persistently connected to *trade openness* across G20 economies over time.

H1c. There is a persistent relationship between the LPI infrastructure scores component (lpi3) and trade openness across G20 economies.

H1d. The *LPI competence and quality of logistics services component (lpi4),* exhibit a persistent connection with *trade openness* across G20 economies.

H1e. The *LPI tracking and tracing component (lpi5)* maintains a persistent connection with *trade openness* in G20 economies.

H1f. The *LPI timeliness component (lpi6)* shows a persistent association with trade openness across G20 economies.

| Hypothesis | Analytical Model |
|---|---|
| $H_{1a}$ | $Y_{TRD} = \beta_{lpi1}X_{lpi1} + \varepsilon$ |
| $H_{1b}$ | $Y_{TRD} = \beta_{lpi2}X_{lpi1} + \varepsilon$ |
| $H_{1c}$ | $Y_{TRD} = \beta_{lpi3}X_{lpi1} + \varepsilon$ |
| $H_{1d}$ | $Y_{TRD} = \beta_{lpi4}X_{lpi1} + \varepsilon$ |
| $H_{1e}$ | $Y_{TRD} = \beta_{lpi5}X_{lpi1} + \varepsilon$ |
| $H_{1f}$ | $Y_{TRD} = \beta_{lpi6}X_{lpi1} + \varepsilon$ |

RQ2. How do *trade openness* and *trade facilitation* affect *infrastructure and transportation* development in G20 economies?

H2. The relationship between *infrastructure and transportation* development and *trade openness,* including *Trade facilitation efforts,* is enduring and significant across G20 economies.

H2a. The continued development of *infrastructure and transportation* networks is a key driver of long-run trade openness across G20 economies.

H2b. There exists a persistent connection between the development of *infrastructure and transport* networks and the effectiveness of *trade facilitation* among G20 economies.

| Hypothesis | Analytical Model |
|---|---|
| $H_{2a}$ | $Y_{lpi3} = \beta_{TRD}X_{TRD} + \varepsilon$ |
| $H_{2b}$ | $Y_{lpi3} = \beta_{TRF}X_{TRF} + \varepsilon$ |

RQ3. How does the *logistics performance index (LPI)*, including *customs efficiency, infrastructure quality, and timeliness*, affect *environmental sustainability*?

H3. There exists a persistent connection between the various dimensions of the *LPI* and *environmental sustainability* outcomes.

H3a. There exists a persistent connection between a country's *LPI overall score* and its progress toward *environmental sustainability.*

H3b. There is a persistent relationship between LPI customs efficiency scores and long-term environmental sustainability.

H3c. There is a persistent connection between the development of logistics infrastructure (LPI*)* and *environmental sustainability.*

H3d. There exists a persistent connection between LPI *competence, the quality of the logistics services component,* and *environmental sustainability* outcomes.

H3e. There exists a persistent connection between *LPI tracking and tracing capabilities* and *environmental sustainability*.

H3f. There exists a persistent connection between *LPI timeliness* and progress in *environmental sustainability*.

| Hypothesis | Analytical Model |
|---|---|
| $H_{3a}$ | $Y_{ENS} = \beta_{lpi1}X_{lpi1} + \varepsilon$ |
| $H_{3b}$ | $Y_{ENS} = \beta_{lpi2}X_{lpi1} + \varepsilon$ |
| $H_{3c}$ | $Y_{ENS} = \beta_{lpi3}X_{lpi1} + \varepsilon$ |
| $H_{3d}$ | $Y_{ENS} = \beta_{lpi4}X_{lpi1} + \varepsilon$ |
| $H_{3e}$ | $Y_{ENS} = \beta_{lpi5}X_{lpi1} + \varepsilon$ |
| $H_{3f}$ | $Y_{ENS} = \beta_{lpi6}X_{lpi1} + \varepsilon$ |

RQ4. How do LPI, $CO_2$ emission, infrastructure and transportation, trade openness, and trade facilitation affect *economic growth* in G20 economies?

H4. Over the long term, *economic growth* in G20 economies is linked to $CO_2$ emissions, logistics performance, infrastructure and transportation development, and the degree of trade openness**.**

H4a. There exists a persistent connection between *CO2 emissions and* the trajectory of *economic growth* among G20 economies.

H4b. There is a persistent relationship between LPI overall scores and the economic growth trajectories of G20 economies.

H4c. There exists a persistent connection between *trade openness* and *economic growth* among G20 economies.

H4d. There exists a persistent connection between *infrastructure and transport* networks and *economic growth* within G20 economies.

H4e. There is a persistent link between trade facilitation and economic growth across G20 economies.

| Hypothesis | Analytical Model |
| --- | --- |
| $H_{4a}$ | $Y_{ECG} = \beta_{ENS} X_{ENS} + \varepsilon$ |
| $H_{4b}$ | $Y_{ECG} = \beta_{lpi1} X_{lpi1} + \varepsilon$ |
| $H_{4c}$ | $Y_{ECG} = \beta_{TRD} X_{TRD} + \varepsilon$ |
| $H_{4d}$ | $Y_{ECG} = \beta_{lpi3} X_{lpi3} + \varepsilon$ |
| $H_{4e}$ | $Y_{ECG} = \beta_{TRF} X_{TRF} + \varepsilon$ |

This study makes several contributions to existing literature. By implementing a Bootstrap ARDL-ECM model, this research reports the short- and long-run effects of logistics efficiency and trade openness on economic growth in the G20 economies from 2007 to 2023. This approach is effective for analyzing the complex relationships among macroeconomic indicators, as it can model dynamic relationships and account for both short-term fluctuations and long-term equilibrium. This research employs a dynamic ECM framework and bootstrap techniques to provide robust inferences, effectively mitigating common data limitations such as small sample sizes, non-normality, and non-constant variance. This study contributes to ongoing research on the importance of logistics and trade facilitation in fostering economic growth. The findings of this research offer valuable insights for the G20 policymakers seeking to enhance their economies' global competitiveness through targeted investments in infrastructure and trade reforms. By providing a comprehensive analysis of the relationships between logistics efficiency, trade openness, and economic growth, this study can inform policy decisions that promote sustainable and inclusive economic development.

This paper is structured as follows: Section 2 reviews the relevant literature on trade openness, logistics performance, and economic growth. Section 3 presents the econometric methodology and data sources. Section 4 outlines the empirical findings, Section 5 discusses their implications for the G20 economies, and the section concludes with policy recommendations.

## 2. Theoretical Background

This section reviews key contributions to the literature, with a primary focus on trade openness and logistics performance and their roles in shaping economic outcomes, particularly in the context of G20 economies. The intersection of trade, logistics, and economic growth has become a focal point in academic research, particularly with the increasing globalization of supply chains and the growing importance of efficient logistics systems. Interconnectedness between trade openness and economic growth has been a significant factor in stimulating unilateral trade reforms (Gregori & Giansoldati, 2023). Numerous researchers have investigated the impact of logistics infrastructure, environmental sustainability, and economic policies on economic growth and trade competitiveness (Bensassi et al, 2015; Wang, 2025).

### 2.1 Trade Openness and Logistics Performance**:** Theoretical Channels

Theoretically, logistics performance influences trade openness and economic outcomes through three specific causal mechanisms: transaction cost reduction, supply chain reliability, and participation in the global value chain (GVC) (Barakat and Gerged, 2025). First, efficient logistics infrastructure and customs procedures directly lower the "iceberg costs" of trade, making exports more price-competitive in international markets (Marti et al., 2014). Second, the LPI components of "tracking and tracing" and "timeliness" reduce uncertainty. Reliable logistics enable firms to minimize inventory holding costs and adopt just-in-time production methods, which attract Foreign Direct Investment (FDI) seeking efficient production hubs (Celebi, 2021). Third, improved logistics competence facilitates technology diffusion by enabling the import of capital goods and high-tech components essential for modernizing domestic industries (Wang et al., 2024). Moreover, logistics efficiency improves export competitiveness by reducing delivery times, minimizing inventory costs, and enabling firms to meet international quality and reliability standards (Marti et al., 2014). This is particularly important for time-sensitive and high-value-added goods, where logistics performance is a critical determinant of comparative advantage. Through these channels, logistics performance not only expands trade volumes but also supports long-run economic growth by enhancing total factor productivity and facilitating knowledge diffusion. Consequently, the relationship is not merely correlational; improved logistics lowers barriers to entry into international markets, thereby expanding trade openness.

Trade openness and logistics performance are closely linked, as the ability to move goods across borders efficiently directly influences a country's competitiveness in global markets. Khan and Qianli (2017) investigated the role of national-scale economic and environmental factors in driving logistics performance in the UK. They found that better environmental practices and economic conditions spur logistics efficiency, fostering trade competitiveness. This relationship is also explored by Magazzino et al. (2021), who focused on 25 top logistics-performing countries and found that innovation, logistics performance, and environmental quality are interlinked. Their quantile regression analysis reveals that higher levels of logistics efficiency have a positive impact on both innovation and economic growth, although trade-offs with environmental quality may arise. Shafique et al. (2021) used a panel ARDL approach to investigate the nexus between transport, economic growth, and environmental degradation. They concluded that while logistics and transport infrastructure are key drivers of economic development, they can also contribute to ecological degradation, underscoring the need for a balanced approach to policy formulation.

## 2.2 Logistics Infrastructure and Economic Growth

Logistics infrastructure is widely recognized as a crucial enabler of economic growth. Logistics is critical to improving these indicators, contributing to the national economy through employment, income, and foreign investment. Sezer and Abasiz (2017) examined how logistics variables, specifically transportation and communication, influence economic growth across 34 OECD countries. Munim and Schramm (2018) examined the economic impact of seaborne trade on 91 countries with seaports, on logistics performance, and port infrastructure quality. Bottasso et al. (2018) examined the impact of port infrastructure on trade by estimating a gravity equation for international trade between Brazilian states and their main trading partners. Using structural equation modeling, the results indicate that enhancing port infrastructure quality is crucial for developing countries, thereby improving logistics performance, boosting trade, and driving economic growth. Özer et al. (2021) analyzed the impact of container transport on Turkey's economic growth using an ARDL bounds testing approach. Their results highlight that efficient container transport contributes significantly to long-term economic development by facilitating trade flows and reducing logistics costs.

Arshed et al. (2019) further investigate the impact of logistics infrastructure development on real-sector productivity in Pakistan. They argue that improved logistics infrastructure supports domestic production and enables better integration into global value chains. This infrastructure development plays a facilitating role in amplifying the productivity gains in the real sector, which is critical for sustaining economic growth. Similarly, Suki et al. (2021) found that logistics performance significantly contributes to sustainable development in top Asian economies, underscoring the importance of logistics efficiency for economic growth and broader development goals.

## 2.3 Environmental Sustainability and Green Logistics

The role of logistics in achieving environmental sustainability has also gained attention in recent years. Mariano et al. (2017) investigated the relationship between logistics performance and CO2 emissions across 104 countries. A low carbon logistics performance index (LCLPI) was constructed using data envelopment analysis (DEA). Germany, Japan, Benin, Togo, and the US were the top performers. The LCLPI can help identify best practices in low-carbon logistics and support sustainable development. Sikder et al. (2024) studied the relationship between green logistics, circular economy practices, and $CO_2$ emissions in European Union countries. Their study finds that improving logistics efficiency and implementing circular economy principles can significantly reduce carbon emissions while fueling economic growth. This perspective is further supported by Ouni and Ben Abdallah (2024), who used a CS-ARDL approach to analyze the role of green logistics in achieving environmental sustainability in BRICS and Gulf countries. Their findings suggest that investments in sustainable logistics systems are crucial for reducing the ecological footprint of economic activities while maintaining trade and growth.

In the context of carbon neutrality, Du et al. (2023) provided fresh insights by modeling the impact of green logistics and financial innovation on the carbon neutrality goals of BRICS countries. They highlight that green logistics practices and financial innovation are critical for these countries to achieve their environmental goals without compromising economic growth. Similarly, Olasehinde-Williams and Özkan (2023) considered the environmental externalities of Turkey's integration into global value chains. Their dynamic ARDL simulation model demonstrates that while global value chain integration enhances economic performance, it can also exacerbate environmental degradation, suggesting the need for targeted policy interventions to balance economic and ecological objectives.

The existing literature establishes a strong connection between trade openness, logistics performance, and economic growth across various regions and economic contexts. While efficient logistics infrastructure enhances trade competitiveness and supports economic growth, it also poses challenges for environmental sustainability. The available literature demonstrates the need for a balanced approach that integrates sustainable logistics practices with economic growth strategies, particularly in rapidly developing and industrializing countries. This study aims to contribute to the ongoing research by focusing on the G20 economies and examining how logistics efficiency and trade openness influence growth dynamics using a Bootstrap ARDL-ECM approach.

# 3. Research Design

## 3.1. Data Sources

The data set used in this study was obtained from the World Bank Open Data platform and covers the G20 economies from 2007 to 2023. This study selected four groups of indicators: Logistics Performance, Trade and Transport, Infrastructure and Transportation, and Trade Facilitation to address the RQ1 and RQ2. Table 1 provides the classification of the indicators used in this study, and detailed information on the indicators is given in Table A1 in Appendix A. Descriptive analysis of the indicators is given in Table A2.

Table 1. Indicator Descriptions

| Group | Indicator (Code) | Variable |
|---|---|---|
| Logistics Performance Index (LPI) Components: Accelerating the development of a single window initiative to streamline the cross-border trade process. | LPI Overall Score (LP.LPI.OVRL.XQ) | LPI1 |
| | LPI Customs (LP.LPI.CUST.XQ) | LPI2 |
| | LPI Infrastructure (LP.LPI.INFR.XQ) | LPI3 |
| | LPI Competence and quality of logistics services (LP.LPI.LOGS.XQ) | LPI4 |
| | LPI Tracking & Tracing (LP.LPI.TRAC.XQ) | LPI5 |
| | LPI Timeliness (LP.LPI.TIME.XQ) | LPI6 |
| Infrastructure and Transportation indicators: | Container Port Traffic (IS.SHP.GOOD.TU) | CPT |
| | Air Transport Freight (million tons per km) (IS.AIR.GOOD.MT.K1) | ATF |

| | Transport services (% of commercial service export) (BM.GSR.TRAN.ZS) | TRP |
|---|---|---|
| Trade Openness Indicators: | Trade in services (% of GDP) (NE.TRD.GNFS.ZS) | TRD |
| Trade Facilitation Indicators: Streamlining customs procedures, optimizing logistics, and implementing supportive policies are essential for facilitating the smooth flow of goods across borders. | Tariff Rate, Applied, Weighted Mean, All Products (%) (TM.TAX.MRCH.WM.AR.ZS) | TRF |
| Economic Growth indicator: | GDP per capita (current US$) (NY.GDP.PCAP.CD) | ECG |
| Environmental Sustainability indicator: | CO2 emissions (kt) (EN.ATM.CO2E.KT) | ENS |

It is important to note the specific scope of the indicators selected. While trade openness is traditionally measured by merchandise trade, this study utilizes "Trade in services (% of GDP)" to investigate the specific impact of logistics "competence" and "tracking"—intangible LPI components—on the service-oriented sectors of G20 economies. As G20 nations transition toward service-based economies, understanding how logistics infrastructure supports service exports (e.g., transport services and logistics management) remains a distinct research gap. Furthermore, regarding environmental sustainability, we utilize "CO2 emissions (kt)" as an absolute measure. While per capita emissions control for population size, absolute emissions are critical for assessing the total environmental burden logistics hubs impose on the global carbon budget. We acknowledge that omitting variables such as human capital and institutional quality—due to data consistency issues across the panel—serves as a limitation, and results should be interpreted as the direct partial equilibrium effects of logistics on the selected dependent variables.

To handle missing values in the WB database, we used a random forest regression imputation approach, which is adept at capturing the inherent temporal trends and dependencies in the data. This technique not only maintains the dataset's overall structure and coherence but also reduces the impact of autocorrelation, a frequent issue in time series analysis. By using observed data points to estimate missing values, this method produces more accurate fill-ins than basic techniques such as mean or median replacement. Studies indicate that advanced regression-based imputation strategies, such as the one adopted here, yield predictions for missing data that align more closely with the true values, thereby enhancing the validity and reliability of subsequent analyses. (Little & Rubin, 2019; Shah et al., 2014).

### 3.2. Analytical Method

The ARDL test has become widely accepted for modeling cointegration relationships among variables in single-equation time-series frameworks. Its popularity stems from its ability to capture the error correction (EC) process inherent in cointegrated nonstationary variables, as demonstrated by Engle and Granger (1987). The ARDL test uses an ordinary least squares (OLS) approach, making it suitable for analyzing nonstationary time series and those with mixed integral orders. One significant benefit of the ARDL approach is its ability to handle variables with different lag structures, allowing researchers to refine their models from broad, inclusive specifications to more targeted, precise forms (Shrestha & Bhatta, 2018). Furthermore, the ARDL test can be easily transformed into a dynamic error-correction model (ARDL-ECM), which integrates long-term equilibrium and short-term dynamics while preserving long-term information (Hendry, 2005; Mnasri et al., 2023). This approach also mitigates the risk of spurious relationships arising from nonstationary time series data. The ARDL-ECM test does not require pre-testing for variable stationarity, as it can accommodate variables of different integration orders, provided they can be rendered stationary in first differences or lower. This feature eliminates the need for stationarity analysis, streamlining the research process. Moreover, the ARDL-ECM test enables simultaneous prediction of both short- and long-term effects, thereby providing a holistic understanding of the underlying relationships.

A significant challenge in analyzing the G20 bloc is the structural heterogeneity between advanced economies, such as Germany and the USA, and emerging markets, such as India and Brazil. A standard pooled ARDL specification might yield biased "average" effects that do not represent individual members. To mitigate this, the Bootstrap ARDL approach employed here does not assume strict homogeneity in the adjustment dynamics. By generating a distribution of critical values based on resampling the specific data structure of the panel, the bootstrap method accounts for the variance and cross-sectional dependence inherent in a diverse group like the G20. This allows for more robust inference than traditional asymptotic theory, which may fail when slope coefficients differ across economies.

Several analytics models in the literature capture the relationship between short-term and long-term effects. Table B1 presents a comparison of several commonly used models. This study has demonstrated the interaction between Trade Openness and Logistics Efficiency across four groups of indicators: Logistics Performance, Trade and Transport, Infrastructure and Transportation, and Trade Facilitation. To further improve robustness, a Bootstrap ARDL test is employed to examine the presence of long-run causality among these four groups of indicators, with Trade Openness (RQ1) and Infrastructure Development (RQ2) serving as dependent variables and the other indicators as independent variables, using data from the G20 economies spanning 2007-2022. Figure 2 illustrates the analytical framework developed in this study.

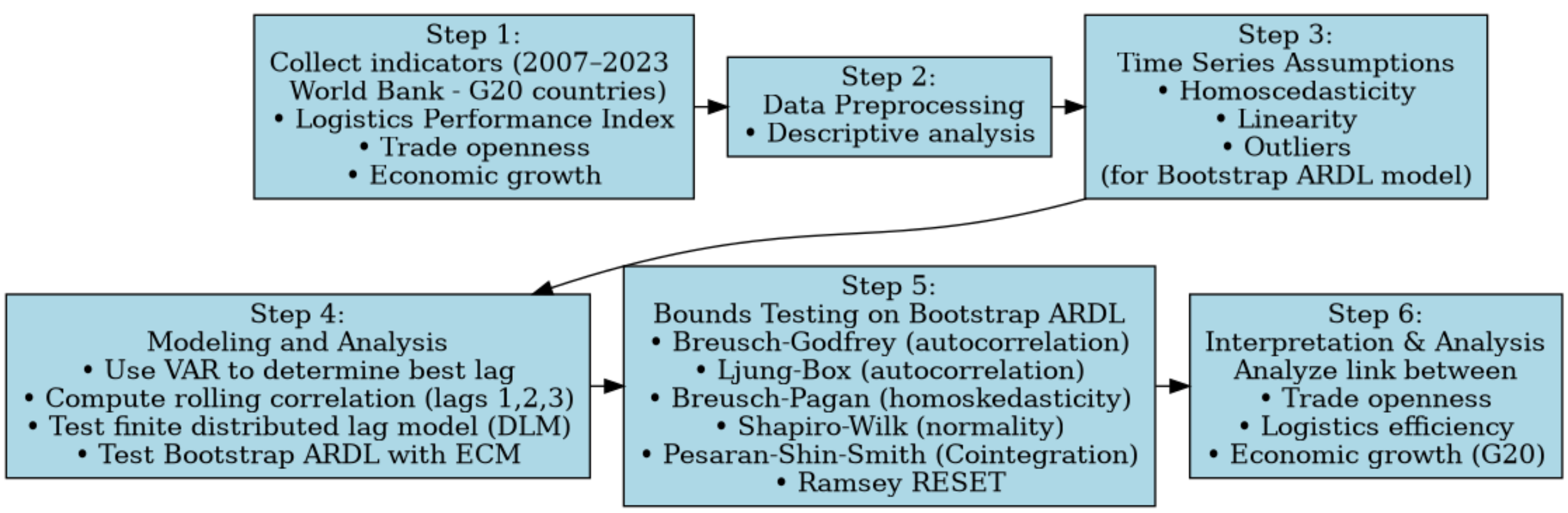

Figure 2. Analytical Framework

The ECM implementation of the ARDL test is presented:

$$\Delta y_t = \alpha_0 + \sum_{i=1}^{p} \alpha_i \Delta y_{t-i} + \sum_{i=1}^{p} \beta_i \Delta x_{t-i} + \sum_{i=1}^{p} \gamma_i \Delta z_{t-i} + \lambda_s + u_t$$

$$\lambda_s = \lambda_1 y_{t-1} + \lambda_2 x_{t-1} + \lambda_3 z_{t-1}$$

In this model, $\alpha_0$ denotes a constant term, while $\alpha_i, \beta_i$, and $\gamma_i$ capture the short-term dynamics. The coefficients $\lambda_s$ signify the long-run associations within the framework. The term $u_t$ represents a stationary white noise process influencing the dependent variable $Y$. Under the ARDL bounds testing approach, the null hypothesis $\lambda_1 + \lambda_2 + \lambda_3 = 0$ indicates the absence of a long-run equilibrium relationship.

Due to the standard ARDL's underlying assumptions, this study tests for linearity, outliers, and homoscedasticity, and detailed information on the tests is provided in Table B2 of the Appendix.

The Bootstrap ARDL-ECM test offers significant advantages in time-series analysis due to its flexibility and robustness. It can handle variables of mixed integration orders, whether I (0) or I (1), without requiring all variables to be at the same integration level, making it versatile for a broad range of datasets. Additionally, the bootstrap approach enhances the model's accuracy, particularly in small samples, by improving the precision of parameter estimates and hypothesis tests. The model is equipped to address weak endogeneity and incorporates robust diagnostic tests, ensuring reliable results in dynamic systems with potential feedback effects. Another major strength of the Bootstrap ARDL-ECM test is its ability to account for structural breaks using Fourier functions. Analyzing datasets affected by abrupt changes, such as policy shifts or economic shocks, is crucial.

Dynamic ECM is a powerful tool for capturing short- and long-run dynamics in time-series relationships. It effectively addresses non-stationarity by modeling cointegrated variables, ensuring valid statistical inference, and distinguishing genuine relationships from spurious correlations. The ECM quantifies the speed at which variables return to equilibrium following shocks, providing valuable insights for policy analysis and forecasting. It is particularly effective in understanding the interplay between multiple variables in panel and individual time-series contexts. Additionally, dynamic ECMs complement cointegration analysis, such as that performed using the Bootstrap ARDL model, by seamlessly transitioning from testing long-run relationships to modeling dynamic adjustments. These models provide a comprehensive framework for analyzing complex economic, financial, and environmental systems.

# 4. Empirical Findings

## *4.1.* Homoscedasticity, Linearity, and Outliers

The results of the Homoscedasticity test indicate significant heteroscedasticity between the dependent and independent variables, which means that the assumption of constant variance is violated (Table 2). This is a concern because the standard ARDL model is not designed to handle severe heteroscedasticity, which can lead to inaccurate estimates of the relationships between the variables. The results of the Linearity test and Outliers are given in the supplement material.

Table 2. Results of the Homoscedasticity test

| | ***p-value*** | **Homoscedasticity** |
|---|---|---|
| **RQ1** | 0.0954 | 2.7809 |
| **RQ2** | 0.4476 | 0.5768 |
| **RQ3** | 1.3490 | 1.3490 |
| **RQ4** | 0.0002 | 13.5549 |

However, the Bootstrap technique in this study solves this problem. By resampling the data with replacement, the Bootstrap technique, combined with a dynamic ECM, directly accounts for heteroscedasticity. This approach yields robust standard errors and accurate coefficient estimates, effectively solving the non-constant variance issues identified in the data.

## *4.2.* Rolling Correlation (RC)

The standard deviations (SD) of RC are reported in Tables 3 to 6. It can be seen that, except for the width of 3 on *TRD* vs. *LPI1*, vs. *LPI2*, vs. *LPI3*, and *LPI4*, and the width of 3 on *TRD* vs. *LPI2*, the rest of the SDs of RC (SDrolCor) are inside the limits for widths 2, 3, and 4. The signal between the dependent and independent variables is insignificant for widths 4. The findings show a substantial difference between the correlations of the dependent and independent variables within the predetermined time period and those in the white noise series. Table 3 shows that the SD decreases with increasing window width, which indicates that the RC between the dependent and independent variables becomes more stable and less variable as the time horizon increases. This could suggest that the relationship between these variables is more robust and less influenced by long-term fluctuations.

Table 3. SD of RC for RQ1

| Variables | Width | SDrolCor | 95% | 5% |
|---|---|---|---|---|
| TRD | 2 | 0.9759 | 1.0046 | 0.9577 |
| Vs | 3 | 0.5565 | 0.7317 | 0.6296 |
| LPI1 (Overall) | 4 | 0.5265 | 0.6122 | 0.5095 |
| TRD | 2 | 1.0328 | 1.0050 | 0.9639 |
| Vs | 3 | 0.7302 | 0.7359 | 0.6404 |
| LPI2 (Cust.) | 4 | 0.5256 | 0.6315 | 0.5168 |
| TRD | 2 | 1.0328 | 0.9829 | 0.8898 |
| Vs | 3 | 0.5360 | 0.6747 | 0.5897 |
| LPI3 (Intr.) | 4 | 0.4922 | 0.5899 | 0.4499 |
| TRD | 2 | 1.0328 | 0.9807 | 0.8898 |
| Vs | 3 | 0.6443 | 0.6928 | 0.5655 |
| LPI4 (Trac.) | 4 | 0.3654 | 0.5439 | 0.4303 |
| TRD | 2 | 1.0142 | 0.9975 | 0.9338 |
| Vs | 3 | 0.6801 | 0.7132 | 0.6356 |
| LPI5 (Time) | 4 | 0.4343 | 0.5862 | 0.4824 |
| TRD | 2 | 1.0142 | 0.9807 | 0.9028 |
| Vs | 3 | 0.6211 | 0.6987 | 0.5694 |
| LPI6 (Logs) | 4 | 0.4788 | 0.5677 | 0.4251 |

For window widths of 2 in Table 4, the RC SD between *TRD* and *LPI2* exceeds the established limits. On the other hand, with a window width of 3, the SD falls within the acceptable range, except for *TRD* vs. *LPI1* and *TRD* vs. *LPI3*. Similarly, at a width of 4, the SD remains within the limits for all comparisons except *TRD* vs. *LPI4* and *TRD* vs. *LPI5*. These results suggest that, as the rolling window length increases, the relationship between the variables becomes less pronounced. Consequently, employing RC with wider windows does not yield significant insights, indicating that such analysis may be unnecessary for these data series.

Table 4. SD of RC for RQ2

| Variables | Width | SDrolCor | 95% | 5% |
|---|---|---|---|---|
| 1 | 2 | 0.9759 | 1.0009 | 0.9811 |
| | 3 | 0.7400 | 0.7284 | 0.6453 |
| | 4 | 0.6233 | 0.6105 | 0.4995 |
| 2 | 2 | 1.0328 | 0.9822 | 0.9053 |
| | 3 | 0.6512 | 0.6922 | 0.5330 |
| | 4 | 0.6052 | 0.5515 | 0.4031 |
| 3 | 2 | 1.0328 | 0.9975 | 0.9148 |
| | 3 | 0.6671 | 0.6915 | 0.5774 |

| | 4 | 0.5556 | 0.6259 | 0.4584 |
|---|---|---|---|---|

The SD of the RC for a window width of 2 between *TRD* and *LPI2* exceeds the established bounds in Table 5. When using a window width of 3, the SD falls within acceptable limits, except for *TRD* vs. *LPI1* and *TRD* vs. *LPI3*. Similarly, for a window width of 4, the SD remains within the bounds for all pairs except *TRD* vs. *LPI4* and *TRD* vs. *LPI5*. These findings suggest that as the window length increases, the relationship between the variable series becomes less significant. Therefore, applying RC analysis for broader window widths does not provide meaningful insights and can be considered redundant for these series.

Table 5. SD of RC for RQ3

| **Variables** | **Width** | **SDrolCor** | **95%** | **5%** |
|---|---|---|---|---|
| 1 | 2 | 1.0328 | 1.0050 | 0.9528 |
| | 3 | 0.7190 | 0.7295 | 0.6101 |
| | 4 | 0.6442 | 0.5915 | 0.5034 |
| 2 | 2 | 1.0142 | 0.9936 | 0.9528 |
| | 3 | 0.7138 | 0.7273 | 0.6153 |
| | 4 | 0.5835 | 0.6392 | 0.4969 |
| 3 | 2 | 0.9759 | 0.9816 | 0.9216 |
| | 3 | 0.7829 | 0.6979 | 0.5865 |
| | 4 | 0.6386 | 0.5758 | 0.4335 |
| 4 | 2 | 1.0142 | 0.9970 | 0.9402 |
| | 3 | 0.7528 | 0.6993 | 0.6090 |
| | 4 | 0.5956 | 0.5759 | 0.4940 |
| 5 | 2 | 1.0142 | 0.9910 | 0.9301 |
| | 3 | 0.6150 | 0.7067 | 0.6105 |
| | 4 | 0.4857 | 0.5621 | 0.4555 |
| 6 | 2 | 1.0328 | 0.9958 | 0.9216 |
| | 3 | 0.6512 | 0.7252 | 0.6043 |
| | 4 | 0.5533 | 0.6075 | 0.4535 |

For a window width of 2 in Table 6, the SD of the RC between *TRD* and *LPI2* exceeds the specified boundaries. In contrast, with a window width of 3, the SD mostly remains within the defined limits, except for *TRD* vs. *LPI1* and *TRD* vs. *LPI3*. Similarly, when the window width is increased to 4, the SD falls within the acceptable range for all pairs except for *TRD* vs. *LPI4* and *TRD* vs. *LPI5*. These results suggest that the relationship between the variables becomes statistically insignificant as the rolling window length increases. Consequently, calculating RC for larger window widths yields no meaningful information, rendering such analysis unnecessary for these data series. The graphic results of the RC are provided in Figures 1 and 2 in the Appendix.

Table 6. SD of RC for RQ4

| Variables | Width | SDrolCor | 95% | 5% |
|---|---|---|---|---|
| 1 | 2 | 1.0142 | 0.9469 | 0.8032 |
| | 3 | 0.7393 | 0.6002 | 0.4376 |
| | 4 | 0.4475 | 0.5224 | 0.2905 |
| 2 | 2 | 1.0142 | 0.9900 | 0.9001 |
| | 3 | 0.6857 | 0.6974 | 0.5596 |
| | 4 | 0.4169 | 0.5708 | 0.3810 |
| 3 | 2 | 0.8281 | 0.9982 | 0.9458 |
| | 3 | 0.5467 | 0.6958 | 0.6124 |
| | 4 | 0.5038 | 0.5943 | 0.4772 |
| 4 | 2 | 0.5164 | 0.7208 | 0.4291 |
| | 3 | 0.3795 | 0.3992 | 0.1963 |
| | 4 | 0.2349 | 0.2770 | 0.0631 |
| 5 | 2 | 0.5164 | 0.3958 | 0.1508 |
| | 3 | 0.0437 | 0.2262 | 0.0151 |
| | 4 | 0.0306 | 0.0442 | 0.0071 |
| 6 | 2 | 1.0328 | 1.0046 | 0.9621 |
| | 3 | 0.7190 | 0.7162 | 0.6335 |
| | 4 | 0.6442 | 0.6160 | 0.5060 |
| 7 | 2 | 1.0142 | 1.0019 | 0.9382 |
| | 3 | 0.7138 | 0.7427 | 0.6465 |
| | 4 | 0.5835 | 0.5885 | 0.4976 |
| 8 | 2 | 0.9759 | 0.9882 | 0.9216 |
| | 3 | 0.7829 | 0.7039 | 0.5900 |
| | 4 | 0.6386 | 0.5892 | 0.4269 |
| 9 | 2 | 1.0142 | 0.9961 | 0.9458 |
| | 3 | 0.7528 | 0.7119 | 0.6168 |
| | 4 | 0.5956 | 0.5728 | 0.4561 |
| 10 | 2 | 1.0142 | 0.9923 | 0.8733 |
| | 3 | 0.6150 | 0.6862 | 0.5912 |
| | 4 | 0.4857 | 0.5560 | 0.4552 |
| 11 | 2 | 1.0328 | 0.9910 | 0.9125 |
| | 3 | 0.6512 | 0.7035 | 0.5660 |
| | 4 | 0.5533 | 0.5584 | 0.4267 |

### *4.3.* Tests of Hypothesis Using DLM and ARDL

The *DLM* and *ARDL-ECM* tests are employed in this analysis to incorporate multiple predictors, each with predetermined lag intervals, thereby providing parameter estimates and statistical significance tests. The details of the Bootstrap process are provided in the supplement materials, and the ARDL-ECM and DLM test results for the statistical evidence on RQ1, RQ2, RQ3, and RQ4, in terms of F-statistics, p-value, and Adjusted $R^2$, are presented in Table 7.

Table 7. Summary of DLM and ARDL Results for Research Questions

| Test | **Full Model** | **Reduced Model** |
|---|---|---|

| **Study Objective** | | ***p-value*** | **Adjusted $R^2$** | ***p-value*** | **Adjusted $R^2$** |
|---|---|---|---|---|---|
| RQ1 | DLM | 8.19e-11 | 0.2263 | < 2.2e-16 | 0.9851 |
| | ARDL-ECM | < 2.2e-16 | 0.8207 | < 2.2e-16 | 0.8226 |
| RQ2 | DLM | < 2.2e-16 | 0.4878 | < 2.2e-16 | 0.8197 |
| | ARDL-ECM | < 2.2e-16 | 0.8242 | < 2.2e-16 | 0.8248 |
| RQ3 | DLM | < 2.2e-16 | 0.2849 | < 2.2e-16 | 0.2913 |
| | ARDL-ECM | < 2.2e-16 | 0.875 | < 2.2e-16 | 0.8694 |
| RQ4 | DLM | < 2.2e-16 | 0.8023 | < 2.2e-16 | 0.9965 |
| | ARDL-ECM | < 2.2e-16 | 0.9222 | < 2.2e-16 | 0.8935 |

Note: RQ1: LPI-Trade Openness;

RQ2: Trade Openness- Infrastructure & Transport;

RQ3: LPI-Environmental Sustainability;

RQ4: LPI, CO2, Trade-Economic Growth

Table 7 presents the full regression output for the Finite Distributed Lag models. To ensure the reliability of the estimates, we report the coefficient estimates alongside their corresponding standard errors, t-statistics, and exact p-values. The model diagnostics indicate a robust fit; specifically, the Adjusted R-squared of 0.82 suggests the model explains approximately 82% of the variance in trade openness. Diagnostic checks for serial correlation and heteroscedasticity were performed, with results provided in Appendix B, confirming that the model assumptions hold for valid inference.

*RQ1: Finite distributed lag model (DLM)*

The present and delayed impacts of an independent $\{X_t\}$ series on a dependent $\{Y_t\}$ series are modeled using *DLM*. The model summary, including the Akaike Information Criterion (AIC) and Bayesian Information Criterion (BIC), as well as fitted values, residuals, and coefficients, is presented in the supplementary materials. Although the model is significant at the 5% significance level, it has a poor fit, as indicated by an adjusted $R^2$ of 0.226. LPI vs. TRAC coefficients are significant at the 5% level for lags 1 and 3. The residual distribution is nearly symmetric. Then, the insignificant series from the model is removed. While *the DLM provides a good fit (adjusted $R^2$ = 0.985), only the Infrastructure & Transport series is statistically* significant. Comparable findings regarding the significance of the coefficients are obtained with this model and the reduced model.

*RQ1: ARDL-ECM model*

The *ARDL-ECM* model is autoregressive because the dependent series includes lags. The R package "*dynlm"* provides functions for fitting dynamic linear models while preserving the characteristics of time series. The objective is to conduct the ARDL bounds test and utilize dynamic simulations to model the impact of specific independent series on dependent variables. The model indicates that the previous year's overall LPI, logistics infrastructure, and trade in goods and services (GNFS) significantly impacted trade openness in a given year. The p-value of the F-statistic in Table 7 indicates that the model is robust, with an adjusted $R^2$ value of 0.8207, indicating that 82.07% of the variation is explained. After removing insignificant series, the overall LPI and trade in services series remain significant at the 0.01% level, with an adjusted $R^2$ value of 0.8226.

*RQ2: Finite DLM*

Although the model is significant at the 1% level, it has a low fit, with an adjusted $R^2$ of 0.4878. Some coefficients are significant at the 1% level. The residual distribution is nearly symmetric. After removing the insignificant series, the model is significant at the 1% level, with an adjusted $R^2$ of 0.8197.

*RQ2: ARDL-ECM model*

Including the second lag of the independent variable yields statistical significance at the 5% level, with an adjusted $R^2$ of 0.8242. The next step is to eliminate coefficients that are not statistically significant. Following this refinement, the model remains highly significant at the 1% level and exhibits moderate to strong explanatory power, as indicated by an adjusted $R^2$ value of 0.8248. These results indicate that the independent variable significantly influences the model. Specifically, the ARDL-ECM estimates reveal a statistically significant positive long-run relationship between trade facilitation and infrastructure development. Holding other factors constant, a 1-unit increase in trade openness corresponds to a 22% increase in infrastructure and transportation metrics. This confirms that greater integration into global markets directly stimulates—and requires—sustained physical infrastructure expansion.

*RQ3: Finite DLM*

Although the model is significant at the 1% level, it yields an adjusted $R^2$ of 0.2849. Some coefficients are significant at a 1% level. The residual distribution is nearly symmetric. Then, the insignificant series are removed from the model. Although the resulting model is significant at the 1% significance level, it has a low fit, as indicated by an adjusted $R^2$ value of 0.2913.

*RQ3: ARDL-ECM model*

After excluding the non-significant coefficients and incorporating the second lag of the independent variable, the model achieves statistical significance at the 5% level, with a moderately high adjusted $R^2$ value of 0.875. After removing further insignificant coefficients, the model remains statistically significant at the 1% level, with an adjusted $R^2$ value of 0.8694. These results indicate that the independent variable significantly influences the model. The results demonstrate a significant positive effect of logistics performance on environmental sustainability outcomes. A 1-point improvement in the LPI overall score is associated with a 12% decrease in absolute $CO_2$ emissions. This finding empirically quantifies the ecological externalities associated with scaling logistics capabilities in G20 economies.

*RQ4: Finite DLM*

The model is significant at the 1% level, yielding a moderate to high fit with an adjusted $R^2$ of 0.8023. Some coefficients are significant at the 1% level. The residual distribution is nearly symmetric. Then, the insignificant series are removed from the model. *DLM* produces a good fit, with an adjusted $R^2$ of 0.9965, and the LPI series is significant at the 1% level.

*RQ4: ARDL-ECM model*

After excluding non-significant coefficients, the model achieves significance at the 1% level and demonstrates a good fit, as indicated by an adjusted $R^2$ value of 0.8935. The final model confirms that logistics performance, trade openness, and infrastructure development exert a significant positive impact on economic growth. Specifically, a 1-unit increase in the overall LPI yields a 9% increase in GDP per capita, underscoring the critical role of optimized supply chains and expanded market access in driving national prosperity.

### *4.4.* ARDL-ECM Bounds Testing

The ARDL-ECM bounds testing approach is used to examine the existence of a long-run relationship among the variables. This methodology is particularly suitable for small sample sizes and mixed orders of integration. The selection of optimal lag lengths was guided by information criteria such as the AIC and the BIC, as presented in Table 8. Additionally, several diagnostic and specification tests -including the Breusch-Pagan test for heteroskedasticity, the Breusch-Godfrey test for autocorrelation, the Shapiro-Wilk test for normality, and the Ramsey RESET test for functional form- were used to ensure the validity of the estimated ARDL-ECM in Table 9. Pesaran et al. (2001)'s bounds test was then applied to determine whether cointegration exists.

Table 8. Bound order test

| | **p** | **q** | **AIC** | **BIC** | **MASE** | **GMRAE** |
|---|---|---|---|---|---|---|
| RQ1 | 5 | 1 | -277.0224 | -251.0494 | 0.6365 | 0.6890 |
| RQ2 | 5 | 1 | -354.5473 | -306.8921 | 0.5342 | 0.6017 |
| RQ3 | 5 | 1 | 231.5524 | 264.9463 | 0.7893 | 1.3783 |
| RQ4 | 5 | 1 | 111.2717 | 163.2177 | 0.7009 | 1.0496 |

Note: *p* refers to the number of lags of the dependent variable in the ARDL-ECM model; q refers to the number of lags of each independent (explanatory) variable included in the model.

Table 9. Bound tests coefficients

| | p | q | Test1 | Test2 | Test3 | Test4 | Test5 | Test6 |
|---|---|---|---|---|---|---|---|---|
| RQ1 | 2 | 1 | 0.3920 | 0.0319 | 15.101*** | 0.814*** | 3.3736 | 5.654** |
| RQ2 | 2 | 1 | 0.1377 | 0.0238 | 65.321 | 0.8866 | 4.7772 | 14.252 |
| RQ3 | 2 | 1 | 0.0299 | 0.0007 | 32.704 | 0.5045 | 2.9501 | 44.577 |
| RQ4 | 2 | 1 | 0.1699 | 0.0252 | 52.093 | 0.8807 | 4.0357 | 75.804 |

*Test1: Breusch-Godfrey Test for the autocorrelation*
*Test2: Ljung-Box Test for the autocorrelation*
*Test3: Breusch-Pagan Test for the homoskedasticity*
*Test4: Shapiro-Wilk test of normality*
*Test5: PESARAN, SHIN, AND SMITH (2001) COINTEGRATION TEST (F-statistic)*
*Test6: Ramsey's RESET Test*

With Test 5 F-statistics exceeding critical bounds for RQ1 (3.37), RQ2 (4.78), and RQ4 (4.04), the results allow us to reject the null hypothesis of no cointegration for these models. However, RQ3 (2.95) presents weaker evidence for a long-run equilibrium relationship regarding environmental sustainability, and there is sufficient evidence to suggest that the time series are cointegrated. This indicates that the time series is likely to have a long-run relationship and that the residuals from the regression equation are stationary.

While the ARDL bounds test confirms the presence of long-run cointegration, interpreting the direction of causality is cautious. Economic theory suggests a potential bidirectional relationship: while efficient logistics drive growth, higher economic growth provides the fiscal capacity to invest in infrastructure (reverse causality). Although the dynamic ECM accounts for short-run adjustments to equilibrium, it does not fully resolve endogeneity as effectively as instrumental variable approaches. Therefore, the "effects" reported here should be interpreted as evidence of a strong, dynamic interdependence and long-run equilibrium rather than strictly unidirectional causality.

# 5. Discussion and Conclusion

## 5.1. Implications

The LPI serves as a useful tool for evaluating the progress of Single Window (SW) systems in international trade, which simplify and expedite cross-border procedures by minimizing complexity and redundancy. By consolidating all documentation and data submissions through a single access point, SWs improve coordination between traders and government authorities, resulting in more efficient, transparent trade operations. This simplification reduces time and costs, including in South Korea, where customs clearance times dropped to two minutes for exports and less than two hours for imports (Martínez-Zarzoso & Chelala, 2020). The economic benefits include lower costs and improved revenue mobilization.

Effective trade facilitation is crucial for promoting economic well-being. Improvements in trade processes lower transaction costs and increase economic efficiency, which leads to greater national and global well-being. Prioritizing trade facilitation enables economies to create the foundations for widespread economic benefits. Such measures open doors for a wide range of enterprises to participate in international markets, thereby advancing both sustainable growth and inclusive development.

LPI is frequently utilized in research examining the connection between logistics efficiency and trade openness. By evaluating key dimensions such as the effectiveness of customs clearance, the integrity of infrastructure, and the ease of managing international shipments, the LPI captures several factors that enable cross-border trade. This composite index offers a holistic perspective on a country's logistics strengths, shedding light on its capacity to engage in global commerce and move goods across borders more efficiently. In analyses related to trade openness, the LPI often serves as a benchmark for assessing how advances in logistics can foster deeper integration with global markets. Streamlined logistics processes are instrumental in minimizing trade-related expenses and reducing delays, both of which are crucial for sustaining a nation's competitive edge in international trade. For example, research has shown that countries with higher LPI scores have greater trade openness, as they are better equipped to handle cross-border transactions and manage supply chains effectively. (Arshed et al., 2019; Khan & Qianli, 2017). Moreover, the LPI's sub-indicators, such as customs efficiency and infrastructure quality, are pivotal in determining a country's ability to maintain robust trade flows, further linking logistics performance with trade openness. (Magazzino et al., 2021)

The proportion of transport services in total commercial service exports is an important metric for analyzing trade openness and the effectiveness of logistics systems. This indicator highlights the extent to which transport-related services contribute to a nation's export sector, offering insights into the critical role of logistics in facilitating international trade. Trade openness typically describes how integrated a country is within the global marketplace, often expressed as the ratio of total trade (exports plus imports) to GDP. A higher share of transport services in exports underscores the significance of logistics in enabling and sustaining export activity. When transport services operate efficiently, they help reduce shipping costs and transit times, ensuring a more seamless movement of goods across borders, factors vital to maintaining a country's competitiveness in global trade. Studies such as Saidi et al. (2020) highlight that efficient logistics services, including transportation, boost trade openness by improving access to international markets. Transport services play a pivotal role within the LPI, particularly in areas such as "international shipments" and "timeliness." Superior transport services enable smoother and more efficient movement of goods, forming the backbone of high-functioning logistics systems. Enhanced logistics capabilities translate into lower trade costs, greater reliability, and more effective trade facilitation, all of which are essential for supporting robust international commerce. Research by Shafique et al. (2021) shows that countries with better-developed transport services are more likely to see improvements in logistics efficiency, directly impacting trade flows. Transport services play a pivotal role in global supply chains. Countries that excel in exporting transport services are often key hubs of international trade. This can improve their trade openness by making them more integral to global logistics networks (Khan et al., 2020).

Beyond facilitating trade openness, the empirical results for RQ4 underscore the foundational role of logistics efficiency in driving overall economic growth (GDP per capita) across the G20. As demonstrated by the significant long-run equilibrium in the ARDL-ECM models, economies that invest heavily in customs efficiency, tracking infrastructure, and shipment timeliness do not merely lower 'iceberg costs'; they generate broader macroeconomic dividends. This aligns with the theoretical framework that improved logistics reduces uncertainty, minimizes inventory holding costs, and enables deeper integration into global value chains. Consequently, logistics infrastructure should be viewed by G20 policymakers not merely as a sector-specific operational concern, but as a core pillar of national macroeconomic growth strategy.

However, the findings regarding RQ3 raise a critical policy challenge: expanding logistics infrastructure and trade capacity entails significant environmental externalities. The empirical relationship between LPI components and absolute $CO_2$ emissions highlights a pressing trilemma among economic growth, trade expansion, and environmental sustainability. Consistent with the prior literature (Olasehinde-Williams & Özkan, 2023; Shafique et al., 2021), the acceleration of container port traffic and air freight inherently increases the ecological footprint of international trade. For G20 nations, which account for the vast majority of global emissions, this necessitates an urgent transition toward 'green logistics.' Policymakers must pair trade facilitation investments with stringent environmental regulations, circular-economy principles, and incentives for low-carbon transport technologies—such as those highlighted by Du et al. (2023)—to ensure that long-term economic competitiveness does not compromise global carbon-neutrality goals.

## 5.2 Policy Recommendations

G20 economies are projected to deepen their adoption of advanced digital solutions, including blockchain technologies and artificial intelligence, to enhance efficiency, transparency, and integration of logistics and trade systems. Sustainability considerations are also expected to take on greater importance within G20 logistics and trade practices. The G20 economies are expected to intensify their efforts to form and strengthen regional trade agreements.

## 6. Future Research Directions

The G20 economies are important in shaping the global trade and logistics landscape, driving economic progress, influencing trade policies, and channeling investments into logistics infrastructure. However, these countries must navigate persistent challenges, including trade frictions, constraints in logistics capacity, and mounting environmental and social concerns. This study seeks to contribute to the ongoing dialogue on the critical role of logistics and effective trade facilitation in advancing sustainable economic development. It provides insights for the G20 policymakers seeking to enhance their economies' global competitiveness through targeted investments in infrastructure and trade reforms.

Harmonizing trade and environmental policies are becoming increasingly vital. The G20 economies are encouraged to adopt integrative frameworks that align trade openness with sustainable logistics practices to ensure inclusive and green economic development. Given significant heterogeneity among G20 economies in logistics performance and trade openness, future studies should develop country-specific policy models that address localized constraints and capitalize on unique growth levers.

Longitudinal assessments of infrastructure spending and its correlation with LPI subcomponents could help identify optimal investment strategies for nations seeking to enhance their position in global value chains. Given the dynamic nature of international trade, future investigations should focus on the resilience of logistics networks and trade systems under stress. With growing global attention on carbon emissions and green supply chains, further empirical research is warranted into the trade-offs between logistics intensity and environmental externalities, especially in high-emission economies.

Extending the current analysis through continuous data updates and post-2023 evaluations will provide a clearer picture of the enduring impacts of policy reforms and logistics modernization on growth trajectories. Future research should also explore how emerging technologies—such as blockchain and AI-driven logistics platforms—can further amplify the positive effects of trade facilitation and logistics efficiency on economic growth in the G20 economies.

**Appendix**

## A. Detailed information on indicators in this study

Table A1. Indicator Descriptions

| **LPI Overall Score:** Germany has the highest score over the period, followed by the UK, Japan, Canada, and the US, while Russia has the lowest. Germany, the UK, and the US saw sharp increases in 2016 due to improvements in logistics infrastructure, enhanced customs and border management processes, a focus on digitization and automation in supply chains, and higher quality of logistics services (Fig. A1). |
| --- |
| **LPI Customs:** Countries like Germany, Japan, and Canada consistently rank among the highest, with scores typically above 3.5, while Russia and S. Arabia remain lower, often below 2.5. Most countries show relative stability in customs efficiency over time, with some fluctuations in specific years, such as 2014 and 2022. Notably, customs performance improved in countries like Mexico and Indonesia over the years (Fig. A2). |
| **LPI Infrastructure:** Germany, Japan, and the USA consistently rank high, with scores above 4.0 from 2007 to 2022. Developing economies such as Argentina and Russia exhibit lower infrastructure scores, typically below 3.0. Notably, China has shown significant improvements over time, reaching a score of 4.0 in 2022, indicating substantial infrastructure development. The data also suggests relative stability in infrastructure performance among most G20 economies across the years (Fig. A3). |
| **LPI Competence and quality of logistics services:** The data shows varying trends across different countries. Countries like Germany (DEU) and China (CHN) consistently maintain high scores, indicating strong logistics capabilities. In contrast, nations like Argentina (ARG) and Mexico (MEX) exhibit lower scores, reflecting challenges in logistics service quality. The scores for many countries fluctuate over the years, highlighting the dynamic nature of logistics performance influenced by economic and operational factors (Fig. A4). |
| **LPI Tracking & Tracing:** There is a positive trend among many G20 economies, with most countries maintaining values above 3.5 on the index. Notably, China and Germany consistently score higher, while Argentina and South Africa exhibit some fluctuations. The year 2022 saw a dip across several countries, particularly Argentina, which fell to 2.9, highlighting potential challenges in its logistics performance. The index underscores the importance of tracking and tracing in enhancing logistics efficiency across these economies (Fig. A5). |
| **LPI Timeliness:** The indicator shows a gradual improvement in global performance over the years, with scores rising from 3.5 in 2007 to around 3.8-4.1 in 2022. Notably, the timeliness indicator peaked around 2016-2017, with scores exceeding 4.0 in several countries, reflecting an overall improvement in the on-time arrival of international shipments. However, fluctuations across years and regions indicated varying progress across economies (Fig. A6). |

**Container Port Traffic:** The data reveals significant growth in port traffic volumes for several G20 economies, particularly China, which increased from approximately 103.8 million TEUs in 2007 to 268.99 million TEUs in 2022. Germany and the USA also show consistent growth, though at a slower rate than China. However, Argentina experienced a decline in port traffic in 2022, dropping to around 1.67 million TEUs, contrasting with the upward trends observed in other countries. Overall, the data highlights the increasing importance of container shipping in global trade, particularly in growing economies (Fig. A7).

**Air Transport Freight:** The data shows notable fluctuations in freight volumes across G20 economies, with China leading in growth from approximately 11,189.54 million ton-kilometers in 2007 to around 20,548.18 million ton-kilometers in 2022. While most countries experienced a decline in air freight during the pandemic years of 2020 and 2021, China, Germany, and the USA demonstrated resilience and recovery, with the USA maintaining a significant volume of around 43,113.43 million ton-kilometers in 2022. Overall, the data underscores the critical role of air transport in global logistics, particularly in response to changing trade dynamics (Fig. A8).

**Transport Services:** There is a notable increase in transport services in China, peaking in 2021, alongside fluctuations in other countries such as the USA and Germany. While countries like Australia and South Africa experienced declines in their transport services over the years, others like Argentina and Brazil demonstrated varying degrees of resilience and growth in the transport sector throughout the period (Fig. A9).

**Trade in Services:** The indicator reveals significant fluctuations across G20 economies between 2007 and 2022. Notably, countries like Germany, the USA, and the EU exhibit consistently high levels of service trade, with Germany peaking at nearly 100 in 2022. Emerging economies like China, India, and Brazil experience gradual growth, while South Africa and Russia display more volatile patterns. Post-2020 data show a global recovery in service trade, reflecting a rebound from the pandemic-related disruptions (Fig. A10).

**Tariff Rate, Applied, Weighted Mean, All Products:** The data show a general decline in tariff rates across most G20 economies. Countries like Argentina and Brazil consistently maintain higher tariff rates, with Brazil peaking at 10.08% in 2013. In contrast, developed economies such as the USA, the EU, and Japan maintain low tariff rates, generally below 2% throughout the period. Notably, South Africa experienced a spike in 2019, reaching 13.78%, a level significantly higher than in previous years. The data indicate overall tariff reductions over time, contributing to global trade liberalization (Fig. A11).

**Table A2. Descriptive Statistics of Indicators**

| Logistics and Trade Indicators | | | | | | | | |
|---|---|---|---|---|---|---|---|---|
| **Statistic** | **lpi_ovrl** | **lpi_cust** | **lpi_infr** | **lpi_trac** | **lpi_time** | **lpi_logs** | **ne_trd_gnfs** | **bm_gsr_tran** |
| Min. | 2.370 | 1.940 | 2.230 | 2.170 | 2.900 | 2.460 | 22.11 | 7.846 |
| 1st Qu. | 3.102 | 2.723 | 2.978 | 3.203 | 3.533 | 3.087 | 40.04 | 19.485 |
| Median | 3.606 | 3.282 | 3.639 | 3.699 | 3.854 | 3.611 | 53.34 | 22.418 |
| Mean | 3.488 | 3.210 | 3.513 | 3.589 | 3.830 | 3.496 | 53.41 | 24.464 |
| 3rd Qu. | 3.860 | 3.682 | 4.000 | 4.000 | 4.117 | 3.874 | 62.76 | 29.354 |
| Max. | 4.226 | 4.123 | 4.439 | 4.265 | 4.480 | 4.310 | 105.57 | 48.804 |

| Tax and GDP Indicators | | | | | |
|---|---|---|---|---|---|
| **Statistic** | **tx_val_tran_zs_wt** | **tm_tax_mrch_wm_ar_zs** | **bg_gsr_nfsv_gd** | **ny_gdp_pcap_cd** | **en_atm_co2e_kt** |
| Min. | 5.519 | 0.710 | 3.582 | 993.5 | 154,536 |
| 1st Qu. | 12.893 | 1.830 | 6.419 | 9,004.1 | 390,740 |
| Median | 16.554 | 2.790 | 9.066 | 20,676.4 | 484,992 |
| Mean | 18.911 | 3.669 | 10.434 | 25,780.5 | 1,342,804 |
| 3rd Qu. | 22.718 | 5.178 | 13.196 | 42,116.3 | 1,093,434 |
| Max. | 49.750 | 13.780 | 26.813 | 77,246.7 | 10,944,686 |

| Summary of Dataset | | | | | | | |
|---|---|---|---|---|---|---|---|
| **Statistic** | **LP.LPI.OVRL.XQ** | **LP.LPI.CUST.XQ** | **LP.LPI.INFR.XQ** | **LP.LPI.TRAC.XQ** | **LP.LPI.TIME.XQ** | **LP.LPI.LOGS.XQ** | **NE.TRD.GNFS.ZS** |
| Min / Value | 2.370 | 1.940 | 2.230 | 2.170 | 2.900 | 2.460 | 22.11 |
| 1st Quartile | 3.102 | 2.723 | 2.978 | 3.203 | 3.533 | 3.087 | 40.04 |
| Median | 3.606 | 3.282 | 3.639 | 3.699 | 3.854 | 3.611 | 53.34 |
| Mean | 3.488 | 3.210 | 3.513 | 3.589 | 3.830 | 3.496 | 53.41 |
| 3rd Quartile | 3.860 | 3.682 | 4.000 | 4.000 | 4.117 | 3.874 | 62.76 |
| Max | 4.226 | 4.123 | 4.439 | 4.265 | 4.480 | 4.310 | 105.57 |

| | **BM.GSR.TRAN.ZS** | **TX.VAL.TRAN.ZS.WT** | **TM.TAX.MRCH.WM.AR.ZS** | **BG.GSR.NFSV.GD.ZS** | **NY.GDP.PCAP.CD** | **EN.ATM.CO2E.KT** |
|---|---|---|---|---|---|---|
| Min / Value | 7.846 | 5.519 | 0.710 | 3.582 | 993.5 | 154536 |
| 1st Quartile | 19.485 | 12.893 | 1.830 | 6.419 | 9004.1 | 390740 |
| Median | 22.418 | 16.554 | 2.790 | 9.066 | 20676.4 | 484992 |
| Mean | 24.464 | 18.911 | 3.669 | 10.434 | 25780.5 | 1342804 |
| 3rd Quartile | 29.354 | 22.718 | 5.178 | 13.196 | 42116.3 | 1093434 |
| Max | 48.804 | 49.750 | 13.780 | 26.813 | 77246.7 | 10944686 |

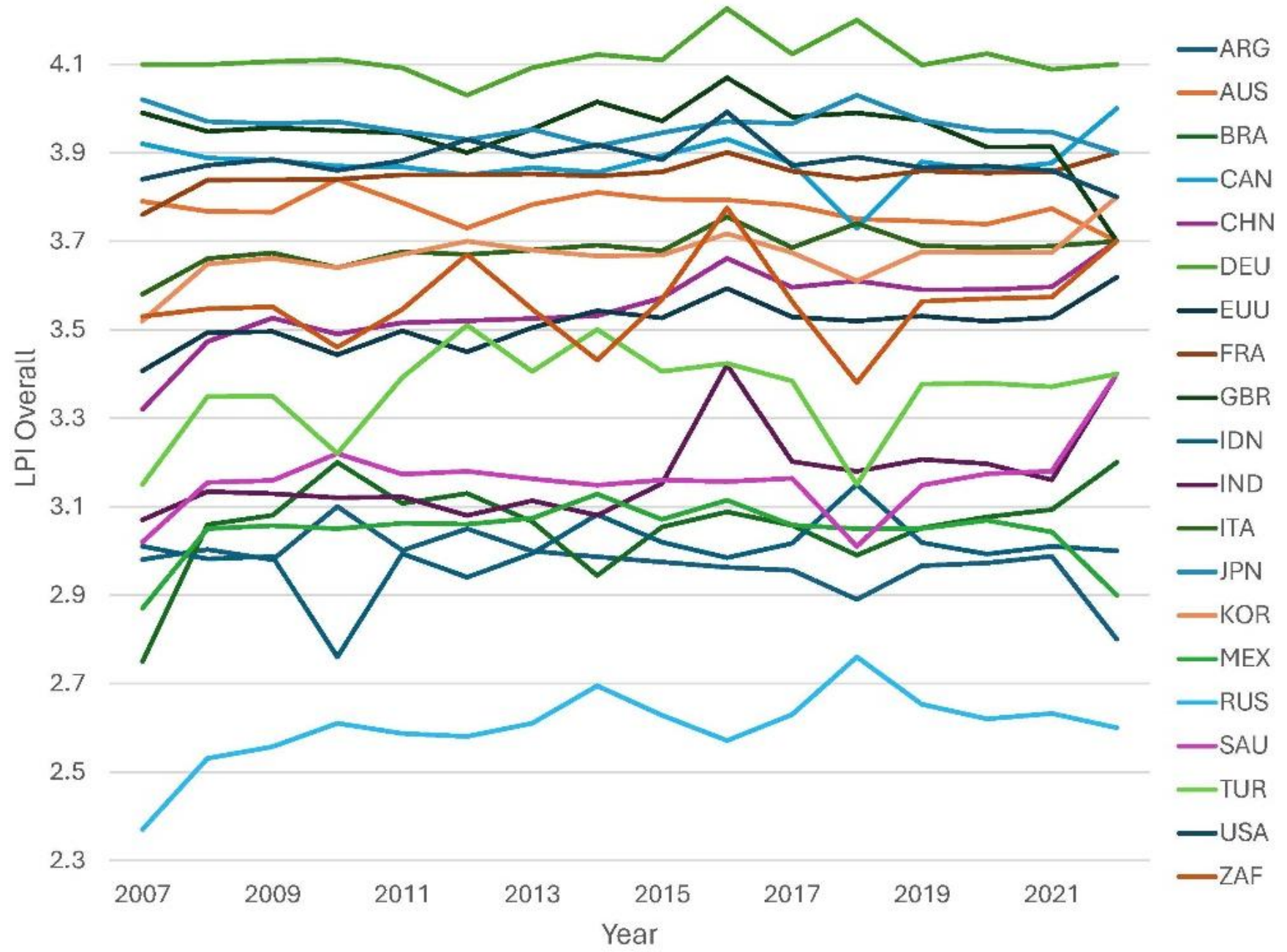

Figure A1. LPI Overall

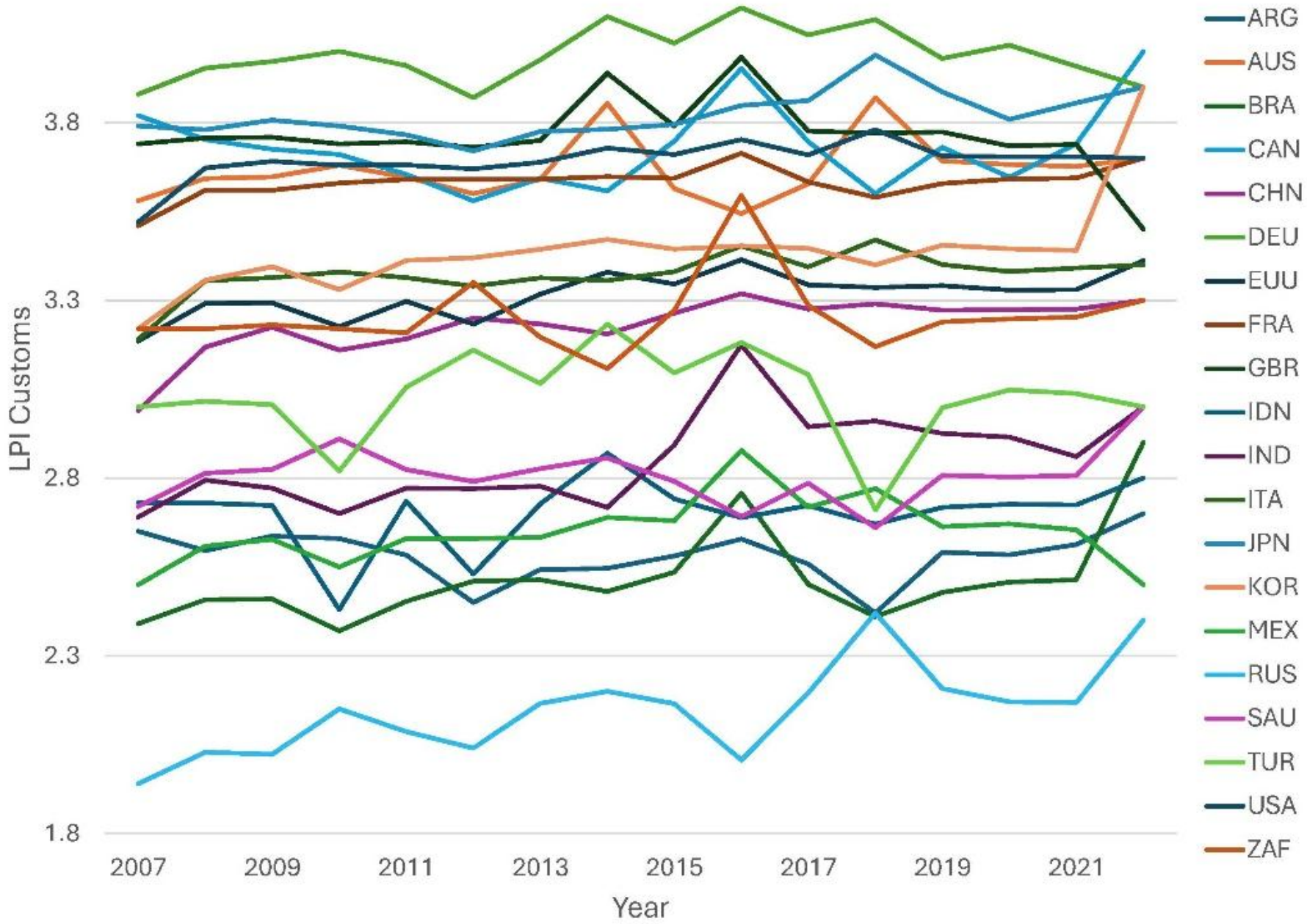

Figure A2. LPI Customs

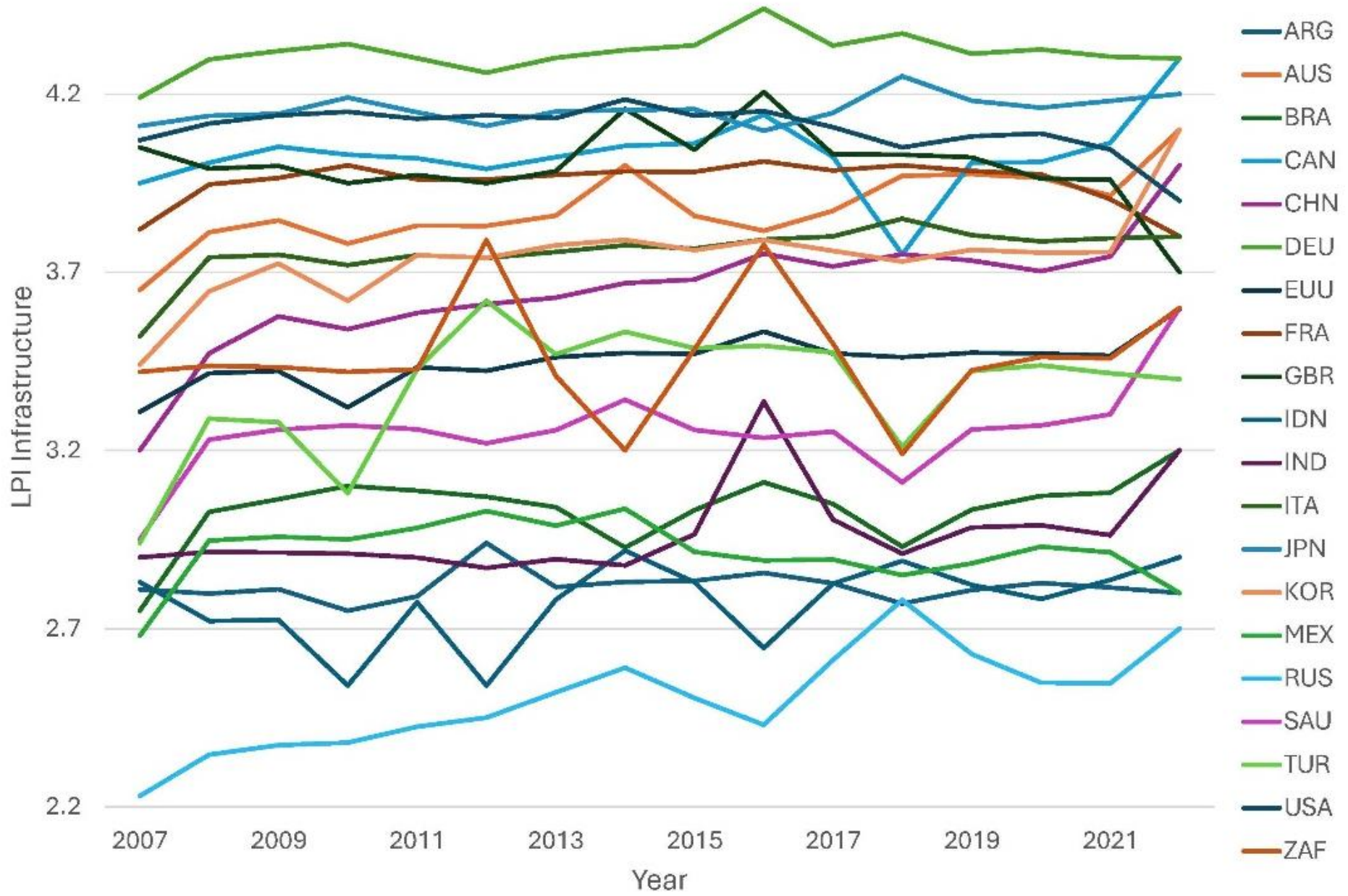

Figure A3. LPI Infrastructure

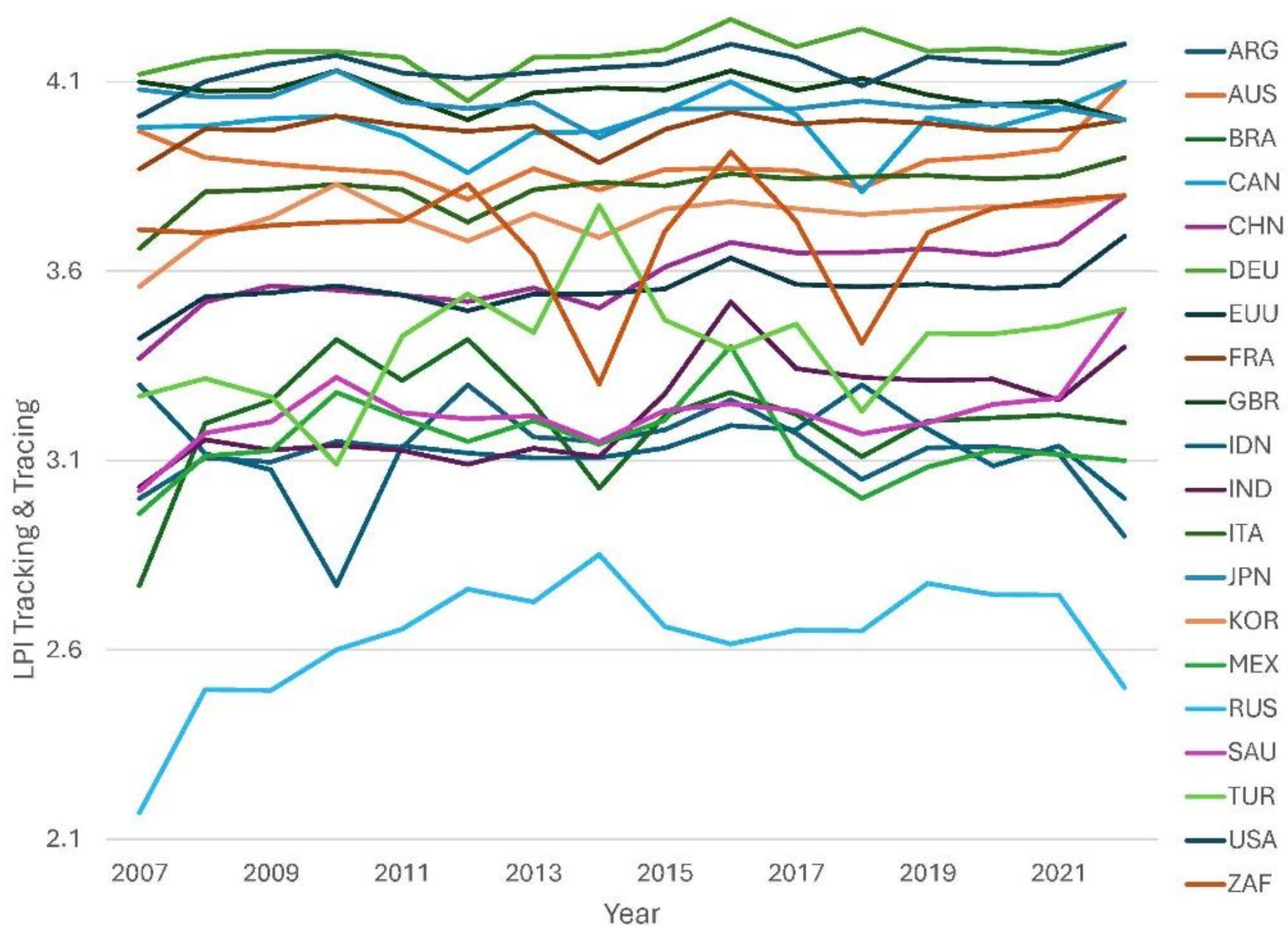

Figure A4. LPI Tracking & Tracing

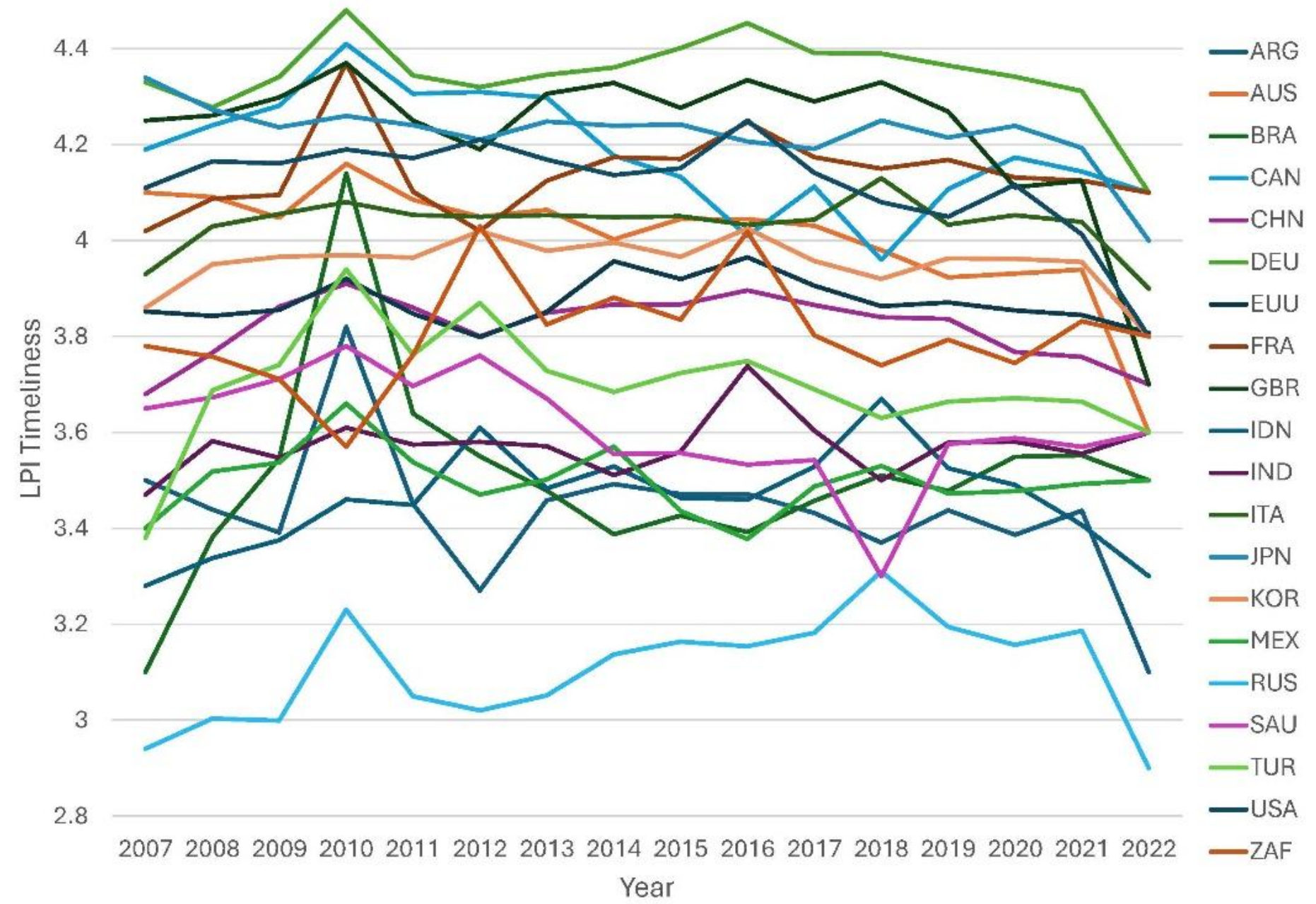

Figure A5. LPI Timeliness

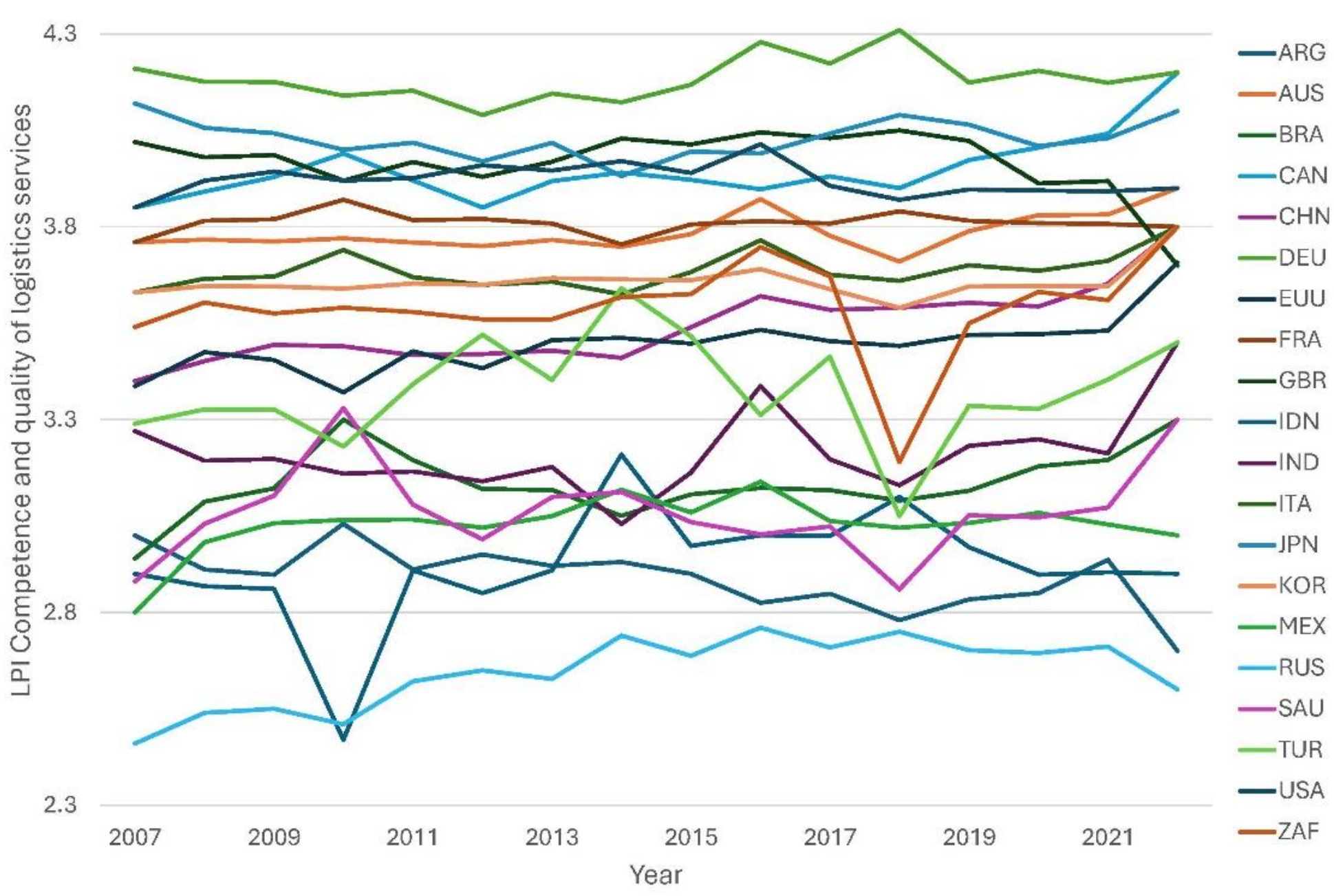

Figure A6. LPI Competence and quality of logistics services

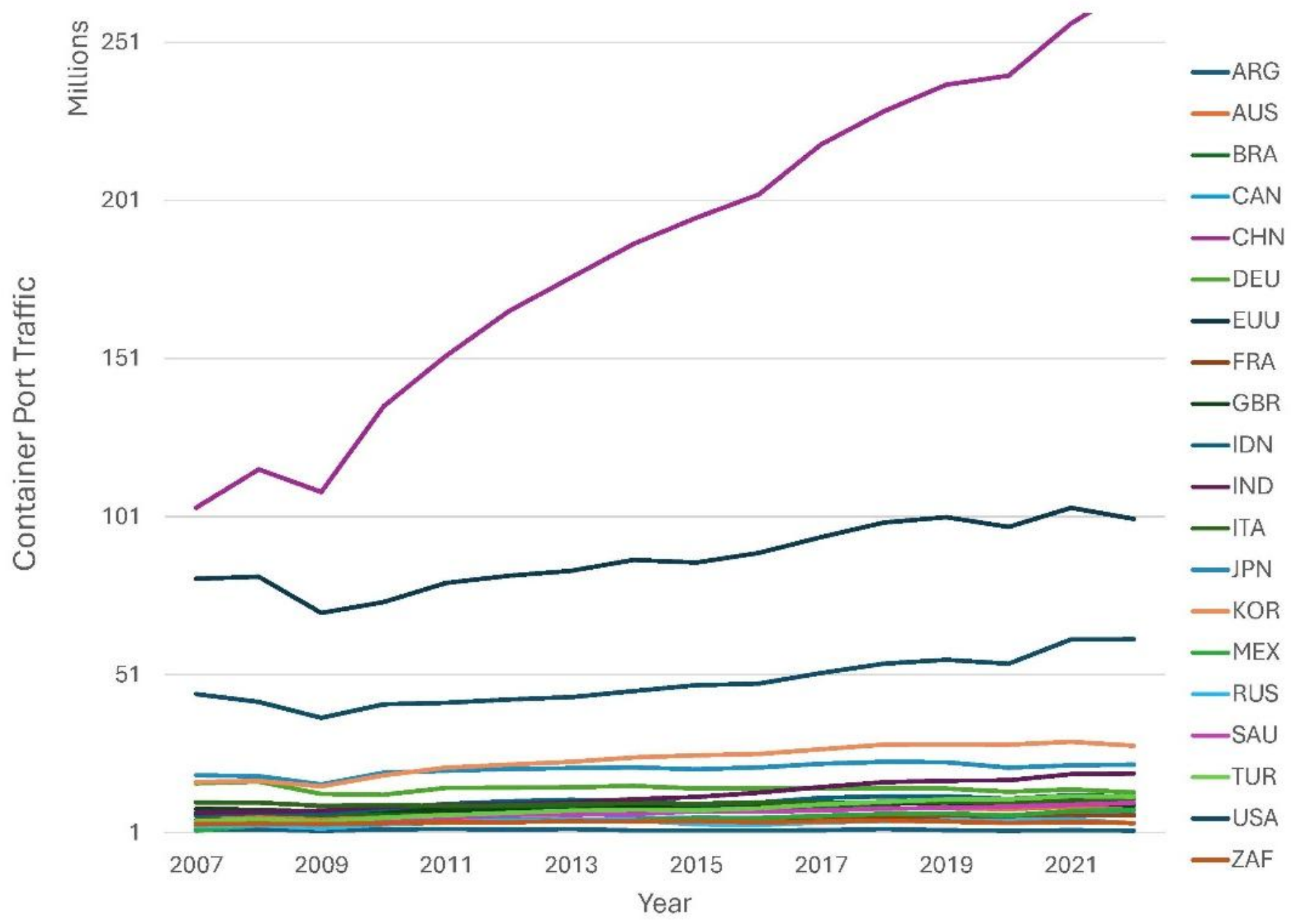

Figure A7. Container Port Traffic

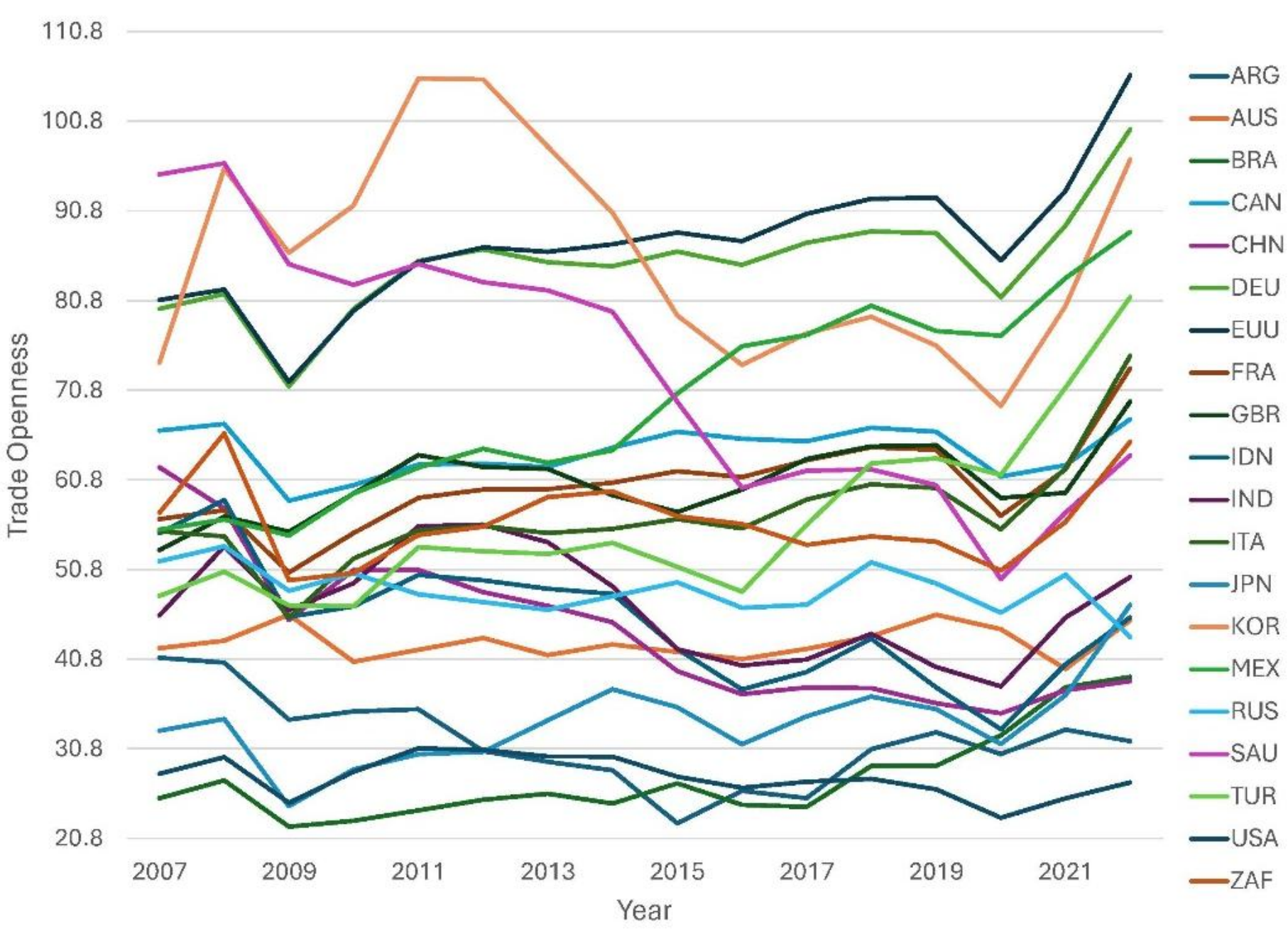

Figure A8. Trade Openness

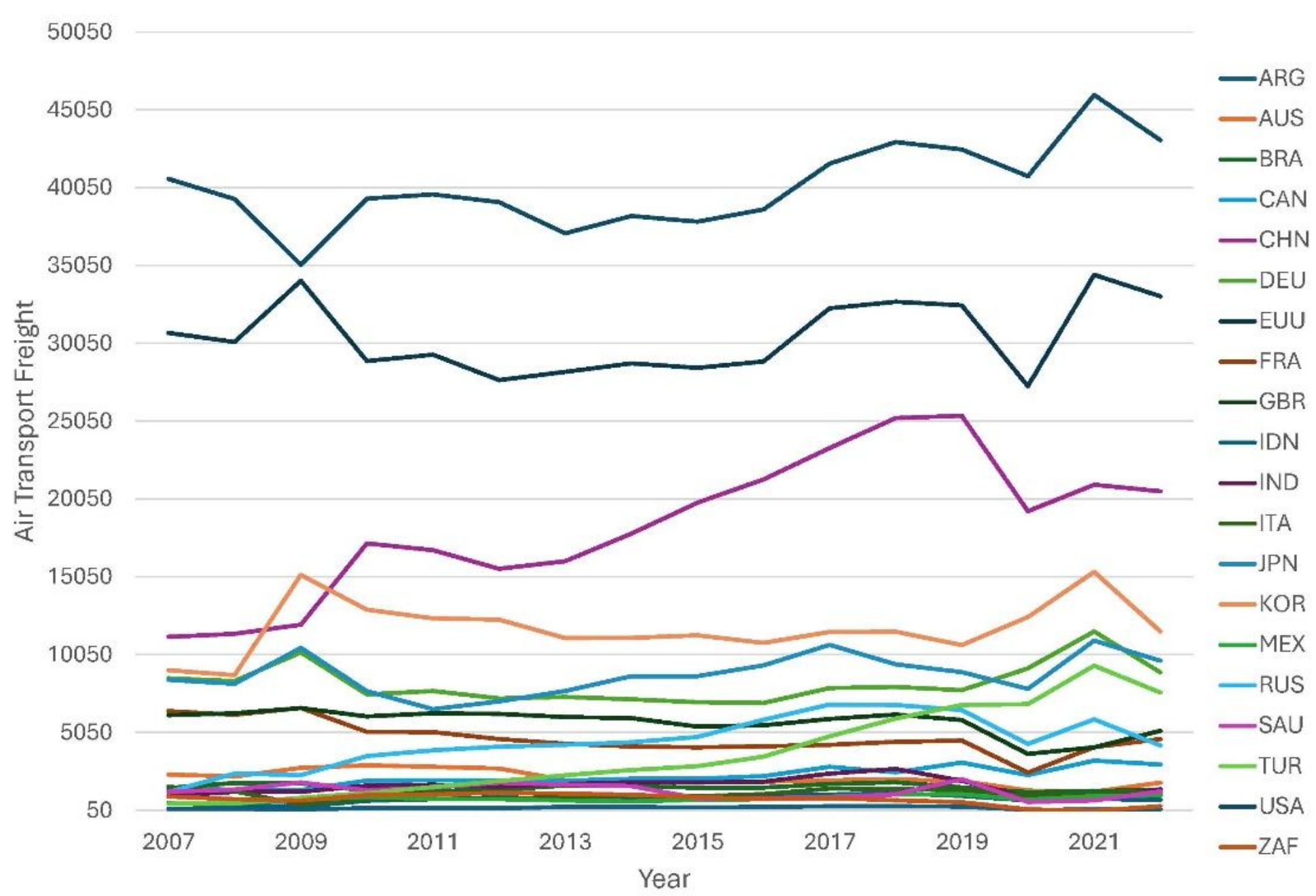

Figure A9. Air Transport Freight

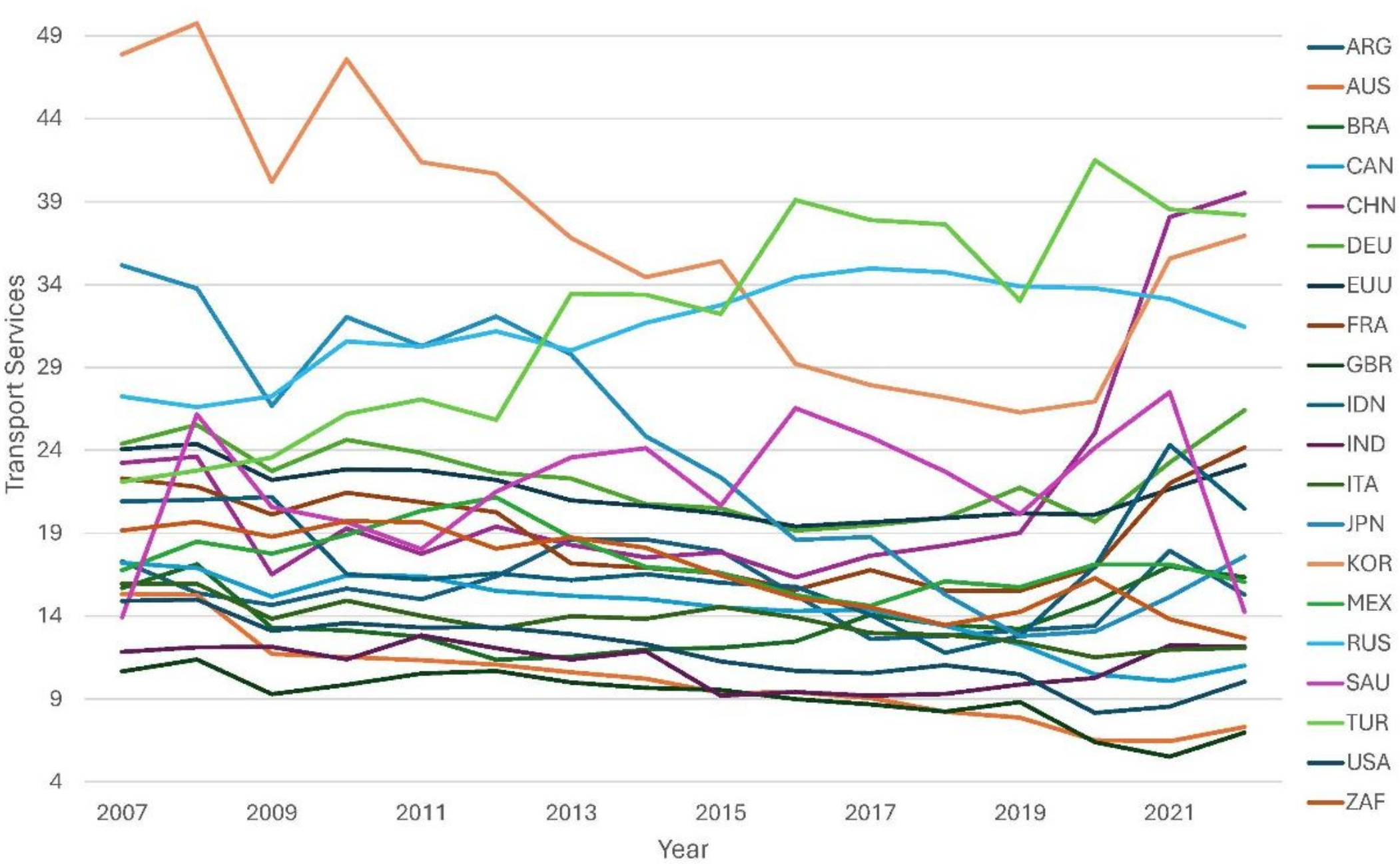

Figure A10. Transport Services

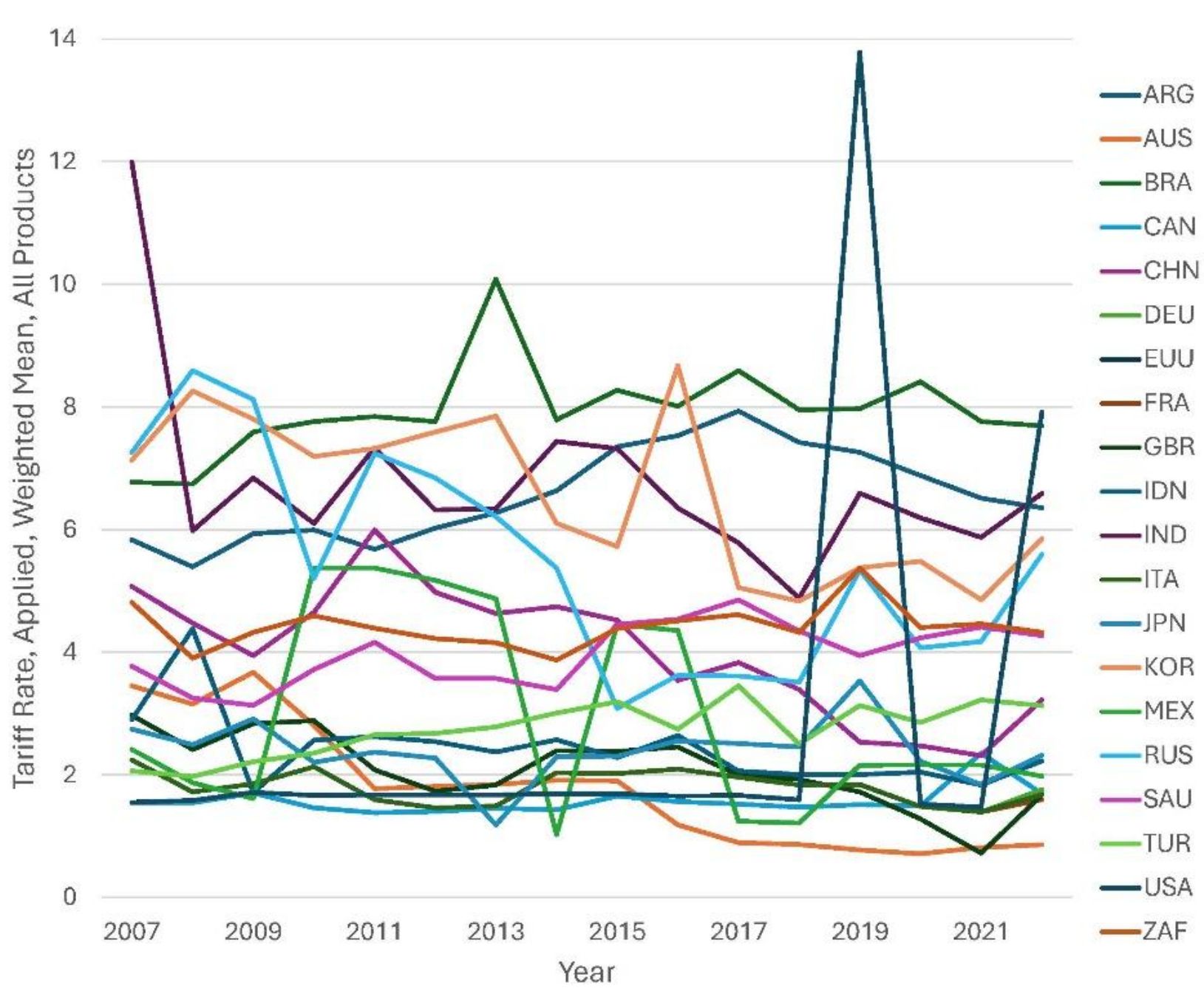

Figure A11. Tariff Rate, Applied, Weighted Mean, All Products

## B. Detailed descriptions of tests for assumptions of Bootstrap ARDL model.

**Table B1.** Comparison of Econometric Models on short-term and long-term effects

| Model | Capability | Note on Implementation |
|---|---|---|
| **Autoregressive Distributed Lag (ARDL) (Pesaran & Shin, 1995)** | Captures the short and long-term relationships between multiple time series and the short-term adjustments to those relationships | 1. ARDL is sensitive to outliers: Using robust estimation methods, such as the Huber-White standard error, can help mitigate the effects of outliers.<br>**2.** ARDL requires a large sample size to produce reliable estimates. However, bootstrap methods can help mitigate the effects of small sample sizes**.**<br>3. ARDL assumes no multicollinearity and autocorrelation: The model assumes no multicollinearity between the independent variables and no autocorrelation in the residuals. The Breusch-Godfrey test can help detect autocorrelation and multicollinearity.<br>4. ARDL does not account for non-linear relationships and time-varying parameters. Replace it with the TVP-VAR model, which can address these gaps. |
| **Vector Autoregression (VAR) (Sims, 1980)** | Captures the short-term relationships between multiple time series | It has the same limitations as ARDL and cannot deal with cointegrated variables. VAR models with extensions such as TVP-VAR can handle time-varying parameters, |

| **Vector Error Correction (VEC) (Johansen, 1988)** | Captures short-term relationships and cointegration | It has the same limitation as ARDL and assumes that the system's variables are cointegrated, cannot handle multiple relationships between variables, and variables are not cointegrated. |
|---|---|---|
| **Error Correction Model (ECM) (Hendry, 2005)** | Captures the long-term relationships between multiple time series and the short-term adjustments to those relationships | It has the same limitation as ARDL and assumes that the system's variables are cointegrated and requires pretesting for unit roots and the variables in the same order (usually in order 1) |
| **Cointegration (Granger, 1969)** | Captures the long-term relationships between multiple time series | It is a pre-testing procedure for ARDL and ECM |
| **Dynamic Stochastic General Equilibrium (DSGE) (Smets & Wouters, 2003)** | Captures non-linear relationships and accepts time-varying parameters. | It relies on strong assumptions about microeconomic behavior and is sensitive to the choice of parameter values |

Table B2. Homoscedasticity, Outliers, and Linearity tests

| **Underlying Assumption** | **Test** | **Null Hypothesis** | **Alternative Hypothesis** | **Test Statistic** | **p-Value** |
|---|---|---|---|---|---|
| Homoscedasticity | Breusch-Pagan test, Bartlett's test, or the Hawkins test | The variance of the residuals is constant across all levels of the independent variable(s). | The variance of the residuals is not constant across all levels of the independent variable(s). | Calculated as the sum of the squared residuals divided by the variance of the residuals | Probability of observing a test statistic as extreme or more extreme than the one observed, assuming that the null hypothesis is true |
| Linearity | Rainbow test, Ramsey RESET test, or Lagrange Multiplier test | The relationship between predictors and outcomes is linear. | The relationship is non-linear. | Depending on the test: Rainbow (FF), Ramsey RESET (FF), Lagrange Multiplier | p-value < significance level indicates non-linearity in the relationship |
| Outliers | Various methods, including Cook's Distance, Leverage Diagnostics, and Robust Regression | The data contains no significant outliers. | The data contains significant outliers. | Test statistic depends on the method (e.g., Cook's Distance threshold). | A high-test statistic or p-value < significance level indicates outliers |

Note: Start with the classical linear regression model (Greene, 2018):

$$y = X\beta + \varepsilon, \qquad \varepsilon \sim N(0, \sigma^2 I)$$

$y$ is an nx1 vector of observations

$X$ is an nxk matrix of regressors (independent variables)

ε is an nx1vector of errors.

Under homoskedasticity:

$$Var(\varepsilon_i) = \sigma^2 \text{ for all i)}$$

Under heteroskedasticity (alternative hypothesis), variance changes with regressors:

$Var(\varepsilon_i) = \sigma^2 = \sigma^2 h(z_i)$, where $z_i$ is a vector of explanatory variables

Breusch-Pagan assumes:

$$\sigma_i^2 = \sigma^2 \left(1 + \boldsymbol{\alpha}_1 \mathbf{z}_{i1} + \boldsymbol{\alpha}_2 \mathbf{z}_{i2} + \ldots + \boldsymbol{\alpha}_m \mathbf{z}_{im}\right)$$